\documentclass[onecolumn,amsmath,amssymb,12pt,superscriptaddress,nofootinbib]{revtex4}

\def\gsim{ \lower .75ex \hbox{$\sim$} \llap{\raise .27ex \hbox{$>$}} }

\def\lsim{ \lower .75ex \hbox{$\sim$} \llap{\raise .27ex \hbox{$<$}} }

\usepackage{graphicx}
\usepackage{dcolumn}
\usepackage{bm}
\input epsf
\def\gsim{ \lower .75ex \hbox{$\sim$} \llap{\raise .27ex \hbox{$>$}} }
\def\lsim{ \lower .75ex \hbox{$\sim$} \llap{\raise .27ex \hbox{$<$}} }

\usepackage{graphicx}
\usepackage{dcolumn}
\usepackage{bm}

\newcommand{\nn}{\nonumber}
\newcommand{\be}{\begin{equation}}
\newcommand{\ee}{\end{equation}}
\newcommand{\bea}{\begin{eqnarray}}
\newcommand{\eea}{\end{eqnarray}}
\renewcommand{\d}{\mathrm{d}}
\def\a{\alpha}

\def\d{\delta}

\def\s{\sigma}
\def\sb{\bar{\sigma}}
\def\e{\epsilon}

\def\i{\iota}

\def\p{\phi}

\def\th{\theta}
\def\tb{\bar{\theta}}

\def\s{\sigma}

\def\pt{\partial}

\def\ad{\dot\alpha}

\def\Db{\bar{D}}
\def\As{A^\ast}
\def\Fs{F^\ast}
\def\P{\Phi}
\def\Pd{\Phi^\dag}

\usepackage{tikz}
\usetikzlibrary{arrows,decorations.markings,mindmap,patterns}

\newlength{\oldoddsidemargin}
\oldoddsidemargin=\oddsidemargin \oddsidemargin=\evensidemargin
\evensidemargin=\oldoddsidemargin

\makeatletter
\def\cleardoublepage{\clearpage\if@twoside \ifodd\c@page\else
   \hbox{}
   \thispagestyle{empty}
   \newpage
   \if@twocolumn\hbox{}\newpage\fi\fi\fi}
\makeatother \clearpage{\pagestyle{plain}\cleardoublepage} 
\usepackage{tabularx}
\usepackage{longtable}
\usepackage{multirow}
\usepackage{graphicx, color}
\usepackage{subfigure}
\usepackage{amsmath}
\usepackage{amsfonts}
\usepackage{amssymb}
\usepackage[english,german]{babel}
\usepackage{bbding}

\usepackage{float}
\usepackage{dsfont}   
\usepackage{hyperref}
\usepackage{mathrsfs} 
\usepackage{makeidx}
\usepackage{listings}
\makeindex
\usepackage{setspace}
\newcolumntype{C}[1]{>{\centering\arraybackslash}b{#1}}
\newcolumntype{D}[1]{>{\centering\arraybackslash}b{#1}}
\newcolumntype{R}[1]{>{\raggedleft\arraybackslash}b{#1}}                                                                                                                                                                                                                                                                                                                                                                                                                                                                                                                                                                                                                                                                                                                                                                                                                                                                                                                                                                                                                                                                                                                                                                                                                                                                                                                                                                                                                                                                                                                                                                                                                                                                                                                                                                                                                                                                                                                                                                                                                                                                                                                                                                                                                                                                                                                                                                                                                                                                                                                                                                                                                                                                                                                                                                                                                                                                                                                                                                                                                                                                                                                                                                                                                                                                                                                                                                                                                                                                                                                                                                                    
\newcolumntype{M}[1]{>{\centering\arraybackslash}m{#1}} 

\def\e{\mathrm{e}}

\def\p{\partial}
\def\d{\mathrm d}

\def\bea{\begin{eqnarray}}
\def\eea{\end{eqnarray}}

\def\S{\mathfrak{S}}
\def\I{\mathrm{i}}

\def\bsm{\left( \!\begin{smallmatrix}}
\def\esm{\end{smallmatrix} \!\right)}

\newcommand{\mac}[1]{\mathcal{#1}}

\def\cE{\mac E}

\def\hD{\hat{D}}

\numberwithin{equation}{section}

\renewcommand{\a}{\alpha}

\renewcommand{\th}{\theta}
\newcommand{\Th}{\Theta}

\newcommand{\veps}{\varepsilon}

\newcommand{\ba}{\begin{align}}
\newcommand{\ea}{\end{align}}

\renewcommand{\i}{\mathrm i}

\newcommand{\lc}[1]{#1\big\vert}

\def\cR{\mac{R}}
\def\cD{\mac{D}}

\def\tcD{\tilde{\cD}}
\def\bcD{\bar{\mac{D}}}
\def\Phid{\Phi^\dagger}
\def\S{\cD\Phi\cD\Phi\bcD\Phid\bcD\Phid}
\def\cDla{{\mac{D}}_{{\alpha}}}

\def\cDua{{\mac{D}}^{{\alpha}}}

\def\bcDla{\bar{\mac{D}}_{\dot{\alpha}}}

\def\bcDua{\bar{\mac{D}}^{\dot{\alpha}}}

\def\cDDa{\cDua\cDla}

\def\bcDDa{\bcDla\bcDua}

\def\weyl{\stackrel{\text{\tiny WEYL}}{\longrightarrow}}
\def\shift{\stackrel{\text{\tiny SHIFT}}{\longrightarrow}}
\def\phone{T}

\def\tbp{\tilde{\bar\psi}}
\def\tp{\tilde\psi}

\def\zf{{\text{0f}}}
\def\of{{\text{1f}}}
\def\tf{{\text{2f}}}

\begin{document}
\selectlanguage{english}
\allowdisplaybreaks
\begin{titlepage}

\title{The Ghost Condensate in $N=1$ Supergravity}

\author{Michael Koehn}
\email[]{michael.koehn@aei.mpg.de}
\author{Jean-Luc Lehners}
\email[]{jlehners@aei.mpg.de}

\affiliation{Max Planck Institute for Gravitational Physics (Albert Einstein Institute), 14476 Potsdam, Germany}
\author{Burt Ovrut}
\email[]{ovrut@elcapitan.hep.upenn.edu}
\affiliation{Department of Physics, University of Pennsylvania,\\ 209 South 33rd Street, Philadelphia, PA 19104-6395, U.S.A.}

\begin{abstract} 

We present the theory of a supersymmetric ghost condensate coupled to $N=1$ supergravity. 
This is accomplished using a general formalism for constructing locally supersymmetric higher-derivative chiral superfield actions.
The theory admits a ghost condensate vacuum in de Sitter spacetime. Expanded around this vacuum, the scalar sector of the theory is shown to be ghost-free with no spatial gradient instabilities. By direct calculation, the fermion sector is found to consist of a massless chiral fermion and a massless gravitino. 
By analyzing the supersymmetry transformations, we find that the chiral fermion transforms inhomogeneously,
indicating that the ghost condensate vacuum spontaneously breaks local supersymmetry with this field as the Goldstone fermion. 
Although potentially able to get a mass through the super-Higgs effect, the vanishing superpotential in the ghost condensate theory renders the gravitino massless. 
Thus local supersymmetry is broken without the super-Higgs effect taking place. This is in agreement with, and gives an explanation for, the direct calculation. 

\vspace{.3in}
\noindent

\end{abstract}
\maketitle

\thispagestyle{empty}

\end{titlepage}

\section{Introduction and Summary}

It was shown in \cite{ArkaniHamed:2003uy} that certain scalar field theories with higher-derivative kinetic terms can, when coupled to gravity, possess a vacuum that is ghost-free but violates the Null Energy Condition (NEC) of general relativity. These ``ghost condensate'' vacua have a number of important applications. For example, ekpyrotic \cite{KOST01,KOST02,Lehners:2008vx} and other bouncing theories \cite{KOSST02,Creminelli:2007aq,Lin:2010pf,Cai:2012va} of the early universe require that spacetime ``bounce'' from a contracting to an expanding phase, perhaps even oscillating cyclically \cite{Steinhardt:2001st,Lehners:2011kr}. From the point of view of low-energy effective field theory, these cosmologies require some form of matter that naturally violates the NEC without introducing any ghosts or singularities in spacetime. 
Such forms of matter are rare--ghost condensates \cite{ArkaniHamed:2003uy} and the closely related Galileons \cite{Nicolis:2008in} are currently the only known scalar field examples. 
Ghost condensates were introduced in this context in new ekpyrotic cosmology \cite{BKO07}. It was shown \cite{BKO07,BKO07b,BKO08} that, given the appropriate potential and kinetic energy functions, the early universe can go through a contracting ekpyrotic phase where a nearly scale-invariant spectrum of scalar perturbations is produced \cite{Lehners:2007ac} (with characteristic non-Gaussian signatures \cite{BKO08,Creminelli:2007aq,Koyama:2007if,Lehners:2007wc,Lehners:2008my,Lehners:2009ja,Lehners:2009qu,Lehners:2010fy}), followed by a ghost-condensate phase where the universe bounces and enters the present epoch of expansion.

All of the above theories involve scalar fields coupled to gravity in the early universe and, hence, it seems essential to understand their ultra-violet (UV) behavior. The quantum divergences of both scalar theories and gravity are under better control within the context of supersymmetry, supergravity and string theory. With this in mind, ghost condensate theories were extended to global supersymmetry in \cite{KLO11}. Specifically, the globally $N=1$ supersymmetric Lagrangian of a single chiral supermultiplet--containing a complex scalar with two real components $\phi$ and $\xi$, a Weyl fermion $\chi$ and an 
auxiliary field $F$--was generalized to include higher-derivative kinetic terms. This theory manifested the ghost condensate vacuum which, due to the appropriate choice of higher-derivative interactions, retained the auxiliary field structure of $F$, was free of spatial gradient instabilities of $\phi$ and had a canonical kinetic energy for the second real scalar $\xi$. The kinetic energy of the fermion evaluated in this condensate vacuum is ghost-free but has a negative spatial gradient term which, perhaps, is physically acceptable. Be that as it may, to resolve this last issue the global supersymmetry construction was extended to more generic higher-derivative interactions in \cite{KLO11b}. This led naturally to a globally $N=1$ supersymmetric theory of conformal Galileons \cite{KLO11b}. Within this context, it was shown that the ghost condensate still persisted with all of the appropriate properties of the original supersymmetric theory but, now, with a correct sign fermion kinetic energy as well. A final, and important, property of the globally supersymmetric ghost condensate vacuum is that it spontaneously breaks supersymmetry. This occurs, not through an expectation value for the $F$ term, but, rather, due to the explicit time-dependence of the scalar condensate. 

These globally supersymmetric condensate theories, although using their eventual interaction with gravity as a motivation for some of their properties, were not a complete coupling to supergravitation. In this paper, we accomplish this by explicitly coupling the higher-derivative chiral superfield Lagrangians introduced in \cite{KLO11} to local $N=1$ supergravity. 
Explicitly we will do the following. After reviewing both scalar and globally supersymmetric ghost condensates in Sections II and III respectively, the basic $N=1$ supergravity ghost condensate Lagrangian is introduced. This is accomplished in Section IV using the general formalism of higher-derivative chiral superfield actions coupled to $N=1$ supergravity introduced in 
\cite{KLO12,KLO12b}. We begin by constructing the Lagrangian, both in superfields and components, for a single chiral superfield with the simplest higher-derivative kinetic term coupled to supergravity. By appropriately choosing the K\"{a}hler potential--in the present paper we do not require a superpotential--solving for the auxiliary fields and Weyl rescaling, the proposed component field supergravity Lagrangian for a ghost condensate theory is presented. This is shown to indeed admit a ghost condensate vacuum in de Sitter spacetime with vanishing gravitino and $\chi$ fermions. The quadratic scalar Lagrangian is evaluated in the condensate vacuum exposing two possible problems--a potential gradient instability in the scalar $\phi$ and an unacceptable kinetic energy for its partner scalar field $\xi,$ which we address later in the paper. 

In the following subsection, the gravitino and $\chi$ kinetic energies and mass terms are presented. By appropriate field redefinition, these are diagonalized and shown to correspond to a massless fermion $\chi$ and a massless gravitino. This result is then interpreted and explained within the context of the fermion supergravity transformations, which are reviewed in Appendix A. As in the global case, the supergravity ghost condensate spontaneously breaks supersymmetry due to the explicit time-dependence of the scalar $\phi$. This renders the supersymmetry transformation of $\chi$ inhomogeneous--signaling the breaking of supersymmetry. However, we show that, as in a Minkowski spacetime vacuum, the gravitino can be redefined so that it transforms homogeneously. Hence, $\chi$ is the massless Goldstone fermion.
The super-Higgs effect
is discussed in detail.
We find that, even though supersymmetry has been spontaneously broken, the gravitino remains massless due to the vanishing of the superpotential and thus the usual super-Higgs effect (by which the gravitino becomes massive) does not take place. These results give an explanation for those obtained by direct diagonalization of the quadratic fermion Lagrangian.
Having understood the fermion kinetic and mass terms, we then return in the next subsection to the problems of the $\phi$ spatial gradient instability and the wrong sign $\xi$ kinetic energy. We present explicit additional supersymmetric terms that, when added to the supergravity Lagrangian, solve both of these problems. Their effect on the ghost condensate vacuum is to make a small shift in the scales of both the condensate and the de Sitter spacetime. The calculation of the requisite component field Lagrangians is presented in detail in Appendix B. Finally, it is shown that these additional terms, while possibly modifying the coefficients of the diagonal gravitino and $\chi$ kinetic energies, still leave the gravitino and $\chi$ as massless fermions. This is accomplished using the generalized fermion transformations presented in Appendix A.

The results of this paper prove the existence of a consistent 
$N=1$ supergravity ghost condensate theory. Although ghost-free, the $\chi$ kinetic energy  continues to manifest a negative spatial gradient term. 
It is of interest, therefore, to extended and generalize the results of this paper to a theory of supersymmetric conformal Galileons coupled to supergravitation--this will appear elsewhere. It is of interest to note that conformal scalar Galileons can occur on the worldvolume of branes \cite{deRham:2010eu} and AdS ``kink'' solitons \cite{KOS12}. Furthermore, it was shown in \cite{OS12} that the bosonic components of $N=1$ supergravity Galileons also appear naturally on the worldvolume of BPS wrapped five-branes in heterotic superstring constructions \cite{LOSW99,LOSW99b,LOW99,DPOR04,BHOP06}. It is tempting to conjecture, therefore, that string soliton worldvolume theories can manifest a ghost condensate solution--thus naturally violating the NEC in a UV complete superstring context. This is presently under study.

\section{Scalar Ghost Condensation}

Let $g_{mn}$ be a $( - + + + )$ Lorentz signature metric of four-dimensional spacetime with coordinates $x^{m}$ and  consider a real scalar field $\phi$. Denote its  standard kinetic term by $X=-\frac{1}{2}(\partial\phi)^{2}$. A ghost condensate vacuum arises from higher-derivative theories of the form
\begin{equation}
{\cal{L}}=\sqrt{-g}P(X) \ ,
\label{a}
\end{equation}
where $P(X)$ is an arbitrary differentiable function of $X$. In a flat Friedmann-Robertson-Walker (FRW) spacetime with metric $ds^{2}=-dt^{2}+a(t)^{2}\delta_{ij}dx^{i}dx^{j}$, and assuming $\phi$ to be dependent on time alone, the scalar equation of motion becomes
\begin{equation}
\frac{d}{dt}\Big(a^{3}P_{,X}\dot{\phi}\Big)=0 \ .
\label{b}
\end{equation}
Clearly this has a trivial solution when $\phi={\rm constant}$. Of more interest is the solution with 
non-constant $\phi$, but for which 
\begin{equation}
X=\frac12 \dot\phi^2 = {\rm constant}, \quad P_{,X}=0 \ .
\label{c}
\end{equation}
Denoting by $X_{\rm ext}$ a constant  extremum of $P(X)$, the equation of motion admits the {\it ghost condensate} solution 
\begin{equation}
\phi=ct \ ,
\label{d}
\end{equation}
where $c^{2}=2X_{\rm ext}$. 

The explicit time-dependence of this solution spontaneously breaks Lorentz invariance and leads to a number of interesting properties. First of all, evaluating the energy and pressure densities one finds 
\begin{equation}
\rho=2XP_{,X}-P  , \quad p=P \quad \Rightarrow \quad \rho+p=2XP_{,X} \ .
\label{e}
\end{equation}
Since by definition $X>0$, it follows that the Null Energy Condition (the NEC corresponds to the requirement $\rho + p \geq 0$) can be violated if $P_{,X}<0$. That is, if we are close to an extremum for $P(X)$, then on one side the NEC is satisfied while on the other it is not. Correspondingly, since Einstein's equations imply $\dot{H}=-\frac12(\rho + p),$ it is now possible to obtain a non-singular bouncing universe--where $H$ increases from negative to positive values. Crucial in determining the viability of this theory is the question of whether or not this NEC-violating solution is ``stable''. To this end, let us expand Lagrangian \eqref{a} to quadratic order in perturbations around the ghost condensate,
\begin{equation}
\phi=ct+\delta\phi(x^{m}) \ .
\label{f}
\end{equation}
We find that
\begin{equation}
\frac{{\cal{L}}}{\sqrt{-g}}=\frac{1}{2}\Big ((2XP_{,XX}+P_{,X})({\dot{\delta \phi}})^{2}-P_{,X}\delta\phi^{,i}\delta\phi_{,i} \Big) \ .
\label{g}
\end{equation}
As a result of Lorentz breaking, the coefficients in front of the time and space pieces are unequal. By inspection, one sees that the condition for the absence of ghosts is that 
\begin{equation}
2XP_{,XX}+P_{,X}>0 \ ,
\label{h}
\end{equation}
which can be achieved around a local minimum
\begin{equation}
P_{,XX}>0
\label{i}
\end{equation}
even in the NEC-violating region where $P_{,X}$ is small but negative. Henceforth, one imposes \eqref{i} in addition to \eqref{c} on the ghost condensate vacuum. This feature is arguably the most striking property of ghost condensate theories, namely, that the NEC can be violated without the appearance of ghosts.

However, in the NEC violating region the coefficient in front of the spatial derivative term in \eqref{g} has the wrong sign. Therefore, the theory suffers from gradient instabilities. These can be softened by adding (small) higher-derivative terms--not of the $P(X)$ type--to the Lagrangian, such as $-(\Box\phi)^{2}$. These modify the dispersion relation for $\delta\phi$ at high momenta and suppress instabilities for a short--but sufficient--period of time. 
In a cosmological context, there are additional constraints arising from a study of the growth of cosmological perturbations, which imply that a non-singular bounce must be fast in order to avoid perturbations from becoming uncontrollably large \cite{Xue:2010ux,Xue:2011nw}.
The bottom line is that bouncing universe solutions via a ghost condensate are admissible, but the bounce is required to occur on a fast time-scale--for more details, see \cite{BKO07}.

\section{Review of Globally $N=1$ Supersymmetric Ghost Condensation}

\subsection*{A. Higher-Derivative Chiral Superfield Lagrangians}

As shown in \cite{KLO11}, the scalar ghost condensate theory can be extended to global $N=1$ supersymmetry. In this paper, we will adopt the notation and conventions of Wess and Bagger \cite{WB92}. A point in flat $N=1$ superspace is labelled by the ordinary spacetime coordinates 
$x^{m}$ and the Grassmann spinor coordinates $\theta^{\alpha}$,${\bar{\theta}}_{\dot{\alpha}}$. One can define superspace derivatives
\be
D_\alpha = \frac{\partial}{\partial\theta^\alpha} +\i \sigma^m_{\a\dot{\a}}\tb^{\dot{\a}} \pt_m\, \qquad \bar{D}_{\dot{\a}} = - \frac{\pt}{\pt \tb^{\dot{\a}}} - \i \th^\a \sigma^m_{\a\dot{\a}}\pt_m
\label{1}
\ee
satisfying the supersymmetry algebra
\be
\{ D_\a , \Db_{\ad} \} = -2\i \s^{m}_{\a\ad}\pt_{m}\,. 
\label{susyalgebra}
\ee
A chiral superfield $\Phi$ is defined by the constraint that 
\be
\bar{D}_{\dot{\a}} \Phi = 0.
\label{2}
\ee
It can be expanded in terms of $\th^{\alpha}$,$\tb_{\dot{\alpha}}$ as
\be
\Phi = A + \i\th\sigma^m\tb A_{,m} + \frac{1}{4}\th\th\tb\tb\Box A + \th\th F + \sqrt{2} \th\chi -\frac{\i}{\sqrt{2}}\th\th \chi_{,m}\sigma^m \tb\,,
\label{3}
\ee
where the component fields are a complex scalar $A(x),$ an auxiliary field $F(x)$ and a spinor $\chi_\a(x)$,  each being functions of the ordinary space-time coordinates $x^m.$ Spinor indices which we do not write out explicitly are understood to be summed according to the convention $\chi \th = \chi^\a \th_\a$ and $\bar{\chi} \tb = \bar\chi_{\dot{\a}} \bar\th^{\dot{\a}}.$ 

The highest ($\th\th\tb\tb$) component of a superfield is automatically invariant under supersymmetry transformations (up to a total spacetime derivative) and, thus, can be used to define a supersymmetric Lagrangian. To isolate the highest component, one can either integrate over the four fermonic coordinates of superspace with the differential $d^4\th \equiv d^2\th d^2\tb,$ or act on the superfield with four superspace derivatives $D^2 \bar{D}^2.$ Both procedures are equivalent. As an example, the ordinary supersymmetric kinetic Lagrangian for chiral superfield \eqref{3} is given by
\be
{\cal{L}}_{\Phi^{\dagger}\Phi}=\int  {\rm d}^4\th\, \Phi^\dag \Phi = \, \Phi^\dag \Phi \mid_{\th\th\tb\tb}=-\pt A \cdot \pt \As + \Fs F + \frac{\i}{2} (\chi_{,m}\s^m \bar\chi - \chi \s^m \bar\chi_{,m})
 \,.
\label{rain1}
\ee
Defining the complex scalar $A$ in terms of two real scalars $\phi$, $\xi$ as
\begin{equation}
A=\frac{1}{\sqrt{2}}(\phi+\i \xi) \ ,
\label{4}
\end{equation}
this Lagrangan becomes
\begin{equation}
{\cal{L}}_{\Phi^{\dagger}\Phi}=-\frac{1}{2}(\partial \phi)^{2}-\frac{1}{2}(\partial \xi)^{2}+ \Fs F + \frac{\i}{2} (\chi_{,m}\s^m \bar\chi - \chi \s^m \bar\chi_{,m}) \ .
\label{5}
\end{equation}
Clearly \eqref{rain1} is the global $N=1$ supersymmetric extension of $X=-\frac{1}{2}(\partial \phi)^{2}$ appearing in the scalar ghost condensate Lagrangian.

To continue, one must provide a global supersymmetric extension of $X^{2}$ as well. This was analyzed in \cite{KLO11} and found, to quadratic order in the spinor $\chi$, to be given by
\bea
&&{\cal{L}}_{D\P D\P \Db \Pd \Db \Pd}=\frac{1}{16}D\P D\P \Db \Pd \Db \Pd\big|_{\th\th\tb\tb} \nn \\
&&~~~\qquad \qquad \quad = (\pt A)^2(\pt\As)^2-2\Fs F \pt A \cdot \pt \As + F^{\ast 2}F^2 \nn \\
&&~~~ \qquad \qquad \quad -\frac{\i}{2} (\chi \s^m\sb^l \s^n \bar\chi_{,n})A_{,m}A^\ast_{,l}+\frac{\i}{2}(\chi_{,n}\s^n\sb^m \s^l \bar\chi)A_{,m}A^\ast_{,l} \nn \\
&&~~~ \qquad \qquad \quad+ \i \chi \s^m \bar\chi^{,n}A_{,m}A^\ast_{,n} -\i \chi^{,m}\s^n \bar\chi A_{,m} A^\ast_{,n}+ \frac{\i}{2} \chi \s^m\bar\chi (A^\ast_{,m} \Box A -A_{,m}\Box \As)\label{6} \\
&&~~~ \qquad \qquad \quad+\frac{1}{2}(F\Box A-\pt F\pt A)\bar\chi\bar\chi + \frac{1}{2}(\Fs\Box \As - \pt \Fs \pt \As)\chi\chi  \nn \\
&&~~~ \qquad \qquad \quad+ \frac{1}{2}F A_{,m}(\bar\chi \sb^m \s^n \bar\chi_{,n}-\bar\chi_{,n}\sb^m \s^n \bar\chi)+\frac{1}{2}\Fs \As_{,m}(\chi_{,n}\s^n \sb^m \chi - \chi \s^n \sb^m \chi_{,n})  \nn \\
&&~~~ \qquad \qquad \quad+\frac{3\i}{2}\Fs F(\chi_{,m}\s^m \bar\chi - \chi \s^m \bar\chi_{,m}) + \frac{\i}{2} \chi \s^m \bar\chi(F \Fs_{,m} - \Fs F_{,m}) \nn \ .
\eea
Written in terms of $\phi$, $\xi$ using \eqref{4}, the pure $A$ term in this Lagrangian becomes
\be
(\pt A)^2 (\pt \As)^2 = \frac{1}{4}(\pt\phi)^4 + \frac{1}{4}(\pt\xi)^4 - \frac{1}{2} (\pt\phi)^2 (\pt\xi)^2 + (\pt\phi\cdot\pt\xi)^2.
\label{new2}
\ee
It follows that \eqref{6} is a gobal $N=1$ supersymmetric extension of the $X^{2}$ term in the scalar ghost condensate Lagrangian. As discussed in \cite{Baumann:2011nm}, there is an alternative supersymmetric extensions of $X^{2}$ given by
\begin{equation}
-\frac{1}{16}(\Phi-\Phi^{\dagger})^{2}\Db D\P D \Db \Pd \ .
\label{7}
\end{equation}
However, \eqref{6} has two properties that render it the appropriate choice. First, it uniquely has the property that when the fermion $\chi$ is set to zero, the only non-vanishing term is the top $\th\th\tb\tb$ component. This makes \eqref{6} useful in constructing higher-derivative terms that include $X^{2}$, a property we will need below. Second, when extended to supergravity--as we will do in the next section--only \eqref{6} leads to minimal coupling of $\phi$, $\xi$ to gravity. The Lagrangian \eqref{7}, on the other hand, produces a derivative interaction $\phi^{2}(\partial \xi)^{2}\cR$ of the chiral scalars with the Ricci scalar $\cR$. 

\subsection*{B. The Supersymmetric Ghost Condensate}

Using \eqref{rain1},\eqref{5} and \eqref{6},\eqref{new2}, one can now present the global $N=1$ supersymmetric extension of the prototypical scalar ghost condensate Lagrangian $P(X)=-X+X^{2},$ with $X_{ext}=\frac12$. 
Since this scalar Lagrangian is purely kinetic with no potential energy, there is no need to consider a superpotential $W.$ This simplifies things, as
\begin{equation}
W=0~~ \Rightarrow~~ F=0
\label{8}
\end{equation}
in the supersymmetric extension. The result, to quadratic order in the fermion $\chi$, is then given by
 \bea
  \nn
 {\cal L}^{\rm SUSY}&=& \Big(-\Phi^\dag \Phi +\frac{1}{16}D\P D\P \Db \Pd \Db \Pd \Big) \Big\vert_{\th\th\tb\tb} \nn \\
 &=& +\frac{1}{2}(\pt\phi)^2 + \frac{1}{4}(\pt\phi)^4 +\frac{1}{2} (\pt\xi)^2 + \frac{1}{4}(\pt\xi)^4 - \frac{1}{2}(\pt\phi)^2(\pt\xi)^2 + (\pt\phi\cdot\pt\xi)^2  \label{gcSusic} \\
 & - &\frac{\i}{2}(\chi_{,m}\sigma^m \bar\chi - \chi \sigma^m \bar\chi_{,m}) -\frac{1}{2}(\pt\phi)^2\frac{\i}{2}(\chi_{,m}\sigma^m \bar\chi - \chi \sigma^m \bar\chi_{,m}) \nn \\
 &-&\phi_{m}\phi_{,n}\frac{\i}{2}(\chi^{,n}\sigma^m \bar\chi - \chi \sigma^m \bar\chi^{,n}). \nn
 \eea
 It was shown in \cite{KLO11} that the associated equations of motion continue to admit a ghost condensate vacuum of the form
 \begin{equation}
\phi=ct~, \quad \xi=0~, \quad \chi=0 
 \label{9}
 \end{equation}
for arbitrary real constant $c$. Recalling, however, that $P_{,X}$ must vanish in a cosmological context, we henceforth restrict to $c=1$.

To assess the stability of the supersymmetric ghost condensate, one can expand in small fluctuations around this background as
\be \phi = t + \delta\phi(t,\vec{x})\,, \qquad \xi = \delta\xi(t,\vec{x})\,, \qquad \chi=\delta\chi(t,\vec{x}) \,.
\label{book1}
\ee
The result, to quadratic order, is
\bea
{\cal L}^{\rm SUSY} &=& (\dot{\delta\phi})^2 + \, 0 \cdot \delta\phi^{,i}\delta\phi_{,i} \nn \\ &+& \, 0 \cdot (\dot{\delta\xi})^2 + \delta\xi^{,i}\delta\xi_{,i} \label{fluctuations} \\ &+& \frac{1}{2} \frac{\i}{2}\left(\delta\chi_{,0}\sigma^0 \delta\bar\chi - \delta\chi \sigma^0 \delta\bar\chi_{,0}\right)  -\frac{1}{2}\frac{\i}{2}\left(\delta\chi_{,i}\sigma^i \delta\bar\chi - \delta\chi \sigma^i \delta\bar\chi_{,i}\right)\,. \nn
\eea
Each line illustrates an important issue to be addressed in supersymmetric ghost condensation.
Note from the first line that $\delta \phi$ has a ghost-free time derivative term, but that the spatial gradient term is vanishing. This reproduces the standard result for a scalar ghost condensate at the minimum of $P(X)$. It follows from the discussion in Section II that $\delta \phi$ will develop a small, negative spatial gradient term in the NEC violating region where $P_{,X}<0$. For the scalar ghost condensate, this is easily cured by including other higher-derivative terms not of the $P(X)$ type--the simplest being $-(\Box \phi)^{2}$. This gradient stabilizing term can be extended to global $N=1$ supersymmetry using the fact, stated above, that when the fermion $\chi$ is set to zero, the only non-vanishing term in $D\P D\P \Db \Pd \Db \Pd $ is the top $\th\th\tb\tb$ component. The appropriate extension was computed in \cite{KLO11} and shown to be
\begin{eqnarray}
&-\frac{1}{2^{11}}D\P D\P \Db \Pd \Db \Pd  \Big(\{D,\bar{D}\}\{D,\bar{D}\}(\Phi + \Phi^\dag)\Big)^2 \Big\vert_{\th\th\tb\tb} = -(\Box\phi)^2\Big(\frac14(\p\phi)^4+\frac{1}{4}(\p\xi)^4\quad~~~~~~~~~~~~ \nn \\
&\quad\quad\quad\quad\quad\quad\qquad\qquad~~~~~~~~~~~~~~~~~~~~~~~~~~+(\p\phi\cdot\p\xi)^2-\frac12(\p\phi)^2(\p\xi)^2\Big) \,,
\label{gradientstabilizer2}
\end{eqnarray}
where we have set $F=0$ and kept only those terms required to analyze the existence and stability of the ghost condensate. We have not displayed terms quadratic in the fermion $\chi$ since each is multiplied by at least one power of $\Box\phi$ and, hence, will vanish in the condensate vacuum.
When this is added to Lagrangian \eqref{gcSusic}, the modified  equations of motion for the component fields continue to admit the ghost condensate solution \eqref{9} with $c=1$. Expanding around this vacuum using \eqref{book1} and $(\partial\phi)^{2}=-1$, we find to quadratic order that \eqref{gradientstabilizer2} becomes
\be
-\frac{1}{4}(\Box \delta\phi)^2\ .
\label{gradientstabilizer3}
\ee
Hence, the first line in the component field Lagrangian is now
\be
{\cal L}^{\rm SUSY} = (\dot{\delta\phi})^2 +  0 \cdot \delta\phi^{,i}\delta\phi_{,i} -\frac{1}{4}(\Box \delta\phi)^2 + \dots \ ,
\label{fluctuations2}
\ee
which softens gradient instabilities by modifying the dispersion relation for $\phi,$ just as in the usual non-supersymmetric ghost condensate \cite{ArkaniHamed:2003uy}. We note that the coefficient in front of the $(\Box\phi)^2$ term has been chosen for convenience here. A wide range of numerical values is in fact possible -- see \cite{BKO07} for a detailed description of the associated bounds.

The second line in \eqref{fluctuations} indicates that the time derivative term in the $\delta \xi$ kinetic energy vanishes, while the spatial gradient term has the wrong sign. This result is new to the supersymmetric extension and, again, must be cured by adding supersymmetric higher-derivative terms. Using the unique properties of $D\P D\P \Db \Pd \Db \Pd$, these were derived in \cite{KLO11} and, to quadratic order in $\xi$, found to be
\bea
&& +\frac{8}{16^2}D\P D\P \Db \Pd \Db \Pd  \Big(\{D,\bar{D}\}(\Phi - \Phi^\dag)\{D,\bar{D}\}(\Phi^\dag - \Phi)\Big) \Big\vert_{\th\th\tb\tb} \nn
\\ &&- \frac{4}{16^3}D\P D\P \Db \Pd \Db \Pd  \Big(\{D,\bar{D}\}(\Phi + \Phi^\dag)\{D,\bar{D}\}(\Phi - \Phi^\dag)\Big)\Big(\{D,\bar{D}\}(\Phi + \Phi^\dag)\{D,\bar{D}\}(\Phi^\dag - \Phi)\Big) \Big\vert_{\th\th\tb\tb} \nn \\
&&=-2(\pt\phi)^4(\pt\xi)^2 - (\pt\phi)^4(\pt\phi\cdot\pt\xi)^2 \,.
 \label{walk1}
\eea
Again, we have displayed only those terms required to analyze the existence and stability of the ghost condensate. When these are added to the Lagrangian, the modified equations of motion continue to admit the ghost condensate vacuum \eqref{9} with $c=1$. Expanding around this vacuum using \eqref{book1}, \eqref{walk1} becomes
\begin{equation}
-2(\partial{\delta \xi})^{2}-(\dot{\delta \xi})^{2} \ .
\label{10}
\end{equation}
Hence, the second line in the component field Lagrangian is now
\begin{equation}
{\cal L}^{\rm SUSY} =\dots +(\dot{\delta \xi})^2 - \delta\xi^{,i}\delta\xi_{,i}+ \dots \ ,
\label{11}
\end{equation}
which is a Lorentz invariant, correct sign kinetic energy for $\delta \xi$.

Finally, consider the $\chi$ kinetic terms in the third line of \eqref{fluctuations}. Although the coefficients are of equal magnitude, the time derivative term is ghost-free while the spatial gradient term has the wrong sign. Using globally supersymmetric extensions of $P(X)$ theories, we have been unable to change the sign of the fermion spatial gradient term while leaving the time derivative term ghost-free. As discussed in \cite{KLO11}, we remain agnostic about whether or not this wrong sign spatial fermion kinetic term is a physical problem. This issue will be further explored within the context of the spontaneous breaking of both global and local supersymmetry. 
It might be worth pointing out though that by extending the ghost condensate model to Galileon theories, the same vacuum solution admits correct-sign, ghost-free fluctuations \cite{KLO11b}.

For completeness, we present the entire globally supersymmetric extension of the ghost condensate theory, combining all of the terms discussed independently above. The result is
\bea
{\cal L}^{\rm SUSY} &=& -\Pd \P \mid_{\th\th\tb\tb} \,  + \frac{1}{16}D\P D \P \Db \Pd \Db \Pd \mid_{\th\th\tb\tb} \nn \\
&& + D\P D \P \Db \Pd \Db \Pd \Bigg[ -\frac{1}{2^{11}}\Big(\{D,\Db\}\{D,\Db\}(\P +\Pd)\Big)^2 \nn \\ && ~\qquad \qquad \qquad \qquad + \frac{1}{2^5}\{D,\Db\}(\P-\Pd)\{D,\Db\}(\Pd-\P) \label{12} \\ && ~\qquad \qquad \qquad \qquad - \frac{1}{2^{10}}\Big(\{D,\Db\}(\P+\Pd)\{D,\Db\}(\P-\Pd)\Big)^2\Bigg] \Bigg\vert_{\th\th\tb\tb}\,. \nn
\eea
In components, writing out all the terms that are relevant for a stability analysis in a ghost condensate background, this corresponds to
\bea
{\cal L}^{\rm SUSY} &=& +\frac{1}{2}(\pt\phi)^2 + \frac{1}{4}(\pt\phi)^4 -\frac{1}{4}(\pt\phi)^4 (\Box \phi)^2 \nn \\ && +\frac{1}{2}(\pt\xi)^2 - \frac{1}{2}(\pt\phi)^2(\pt\xi)^2 - 2 (\pt\phi)^4 (\pt\xi)^2 + (\pt\phi \cdot\pt\xi)^2 - (\pt\phi)^4 (\pt\phi\cdot\pt\xi)^2 \label{13} \\ && +\frac{\i}{2}(\chi_{,m}\s^m \bar\chi - \chi \s^m \bar\chi_{,m})\Big(-1-\frac{1}{2}(\pt\phi)^2\Big) -\phi_{m}\phi_{,n}\frac{\i}{2}(\chi^{,n}\s^m \bar\chi - \chi \s^m \bar\chi^{,n}) \ . \nn
\eea

\subsection*{C. A New Form of Supersymmetry Breaking}

The supersymmetric ghost condensate manifests another important property. Consider the supersymmetry transformation of the spinor,
\bea \delta \chi &=& \i \sqrt{2} \sigma^m \bar\zeta \pt_m A + \sqrt{2} \zeta F\,.
\label{14}
\eea
Ordinarily, spontaneous breaking of supersymmetry is achieved by having a non-zero, constant vev of the dimension-two auxiliary field $F$, thus rendering the transformation inhomogeneous.  The spinor $\chi$ then becomes the Goldstone fermion of the spontaneously broken supersymmetry.

With the supersymmetric ghost condensate, we find ourselves in a new situation. In this vacuum, the vev of $F$ vanishes.
Now, however, supersymmetry is broken by the scalar field $A$ getting a non-zero and, moreover, linearly time-dependent vev $\langle \dot{A} \rangle = \langle \dot{\phi} \rangle/\sqrt{2}=c/\sqrt{2}$, where we restore the arbitrary dimension-two constant. Therefore,
\bea
\delta \chi &=& \i \sqrt{2} \sigma^m \bar\zeta \pt_m A =\i \s^0 \bar\zeta c \ .
\label{15}
\eea
As previously, the fermion transforms inhomogeneously and, hence, supersymmetry is spontaneously broken. For the ghost condensate, however, the inhomogeneous term arises from the linear time-dependent vev of $\phi$ rather than from the $F$-term.
The scale of supersymmetry breaking corresponds to the scale of the ghost condensate.
It is of interest to explore this mechanism within the context of supergravity. There, one might expect the Goldstone fermion to be eaten by the gravitino, and to render the latter massive. However, because of the wrong-sign spatial kinetic term of the spinor and other properties of the ghost condensate background--as discussed in the previous subsection--there may well be subtleties involved. We will return to this intriguing question in the next section.

\section{The Ghost Condensate in $N=1$ Supergravity}

In this section, we couple the globally supersymmetric Lagrangian given in \eqref{12} to $N=1$ supergravity and discuss the ghost condensate vacuum in this context.  As above, only those terms in the component field Lagrangian that have up to two fermions are presented--since this is all that is required to discuss the supergravity ghost condensate.

\subsection*{A. The Chiral Superfield Lagrangian in Supergravity}

In \cite{KLO12}, it was shown that a global $N=1$ supersymmetric Lagrangian of the form
\begin{eqnarray}
&&{\cal L}^{\rm SUSY} = K(\Phi,\Phi^{\dagger}) \mid_{\th\th\tb\tb} \,  + \frac{1}{16}D\P D \P \Db \Pd \Db \Pd T(\Phi, \Phi^{\dagger},\partial_{m}\Phi, \partial_{n}\Phi^{\dagger})\mid_{\th\th\tb\tb} \nn \\
&&~~\quad \qquad +\Big(W(\Phi) \mid_{\th\th}+W^{\dagger}(\Phi^{\dagger}) \mid_{\tb\tb}\Big) \ ,
\label{16}
\end{eqnarray}
where $K$ is any real function of $\Phi,\Phi^{\dagger}$, $T$ is an arbitrary hermitian function of $\Phi,\Phi^{\dagger}$ with any number of their spacetime derivatives (with all derivative indices contracted) and $W$ is an arbitrary holomorphic function of $\Phi$, can be consistently coupled to supergravity\footnote{Related work of interest includes \cite{Buchbinder:1988yu,Buchbinder:1994iw,Banin:2006db,Brandt:1993vd,Brandt:1996au,Antoniadis:2007xc,Farakos:2012qu}.}. 
This was accomplished within the context of curved superspace, following the notation and formalism introduced in \cite{WB92}. Suffice it here to say that a point in curved $N=1$ superspace is labelled by $(x^{m}, \Theta^{\alpha},{\bar{\Theta}}_{{\dot{\alpha}}})$ and that the chiral projector is $\bcD^2 - 8R,$ where $\bcD_{\dot\a}$ is a spinorial component of the curved superspace covariant derivative $\cD_A=(\cD_a,\cD_\a,\bcD_{\dot\a})$ and $R$ is the curvature 
superfield\footnote{All covariant derivatives used in this paper contain the superspin connection {\it only}. The $U(1)$ connection associated with K\"{a}hler transformations--sometimes absorbed into the covariant derivatives in \cite{WB92}--are, in this paper, always written out explicitly.}. In its component expansion, $R$ contains the Ricci scalar $\cR$ and the gravitino $\psi_m,$ as well as the auxiliary fields of 
supergravity--namely a complex scalar $M$ and a real vector $b_m.$ The components in the $\Th$ expansion of $R$ are
\begin{align}
R & = -\frac16 M-\frac16\Theta^\alpha\big({\sigma_{\alpha\dot\alpha}}^a\bar\sigma^{b\dot\alpha\beta}\psi_{ab\beta}-\I{\sigma_{\alpha\dot\alpha}}^a\bar\psi_a{}^{\dot\alpha}M+\I\psi_{a\alpha}b^a\big) \notag\\
&\quad+ \Theta^\alpha\Theta_\alpha \Big(\frac1{12}\cR-\frac16\I\bar\psi^a{}_{\dot\alpha}\bar\sigma^{b\dot\alpha\beta}\psi_{ab\beta}-\frac19MM^*-\frac1{18}b^ab_a+\frac16\I {e_a}^m\cD_m b^a \label{17}\\
&\qquad\qquad\quad-\frac1{12}\bar\psi_{\dot\alpha}\bar\psi^{\dot\alpha}M+\frac1{12}{\psi_a}^\alpha{\sigma_{\alpha\dot\alpha}}^a\bar\psi_c{}^{\dot\alpha}b^c-\frac1{48}\veps^{abcd}\left[\bar\psi_{a\dot\alpha}\bar\sigma_b{}^{\dot\alpha\beta}\psi_{cd\beta}+{\psi_a}^\alpha\sigma_{\alpha\dot\alpha b}\bar\psi_{cd}{}^{\dot\alpha}\right]\Big) \ . \nn 
\end{align}
A second superfield that we will need is the chiral density $\mac{E}$. This contains the determinant of the vierbein $e_{m}^{a}$, as well as $M$ and $\psi_m.$ Its component expansion is 
\be
2\cE = e\Big(1+\I\Theta^\alpha{\sigma_{\alpha\dot\alpha}}^a{\bar\psi_a}{}^{\dot\alpha}-\Theta^\alpha\Theta_\alpha \left[M^*+\bar\psi_{a\dot\alpha}\bar\sigma^{ab\dot\alpha}{}_{\dot\beta}\bar\psi_b{}^{\dot\beta}\right]\big)\ .
\label{18}
\ee
In terms of these quantities, the $N=1$ supergravity extension of Lagrangian \eqref{16} is
\begin{eqnarray}
&&\mac{L}^{SUGRA} = \int \d^2\Theta 2\mac{E}\Big[\frac38(\bcD^2-8R)\e^{-K/3}-\frac18(\bcD^2-8R)(\S T) \nn \\
&&\qquad\qquad\qquad\qquad\qquad+W(\Phi)\Big]+{\rm h.c.} 
\label{19}
\end{eqnarray}
Partially expanded in component fields, this becomes
\begin{eqnarray}
&&\mac{L}^{SUGRA} =\Big[-\frac3{32}e(\cD^2\bcD^2\e^{-K/3})+\I\frac3{16}e\bar\psi_{a\dot\alpha}\bar\sigma^{a\dot\alpha\alpha}(\cDla\bcD^2\e^{-K/3})-\frac38e\big(M^*+\bar\psi_a\bar\sigma^{ab}\bar\psi_b\big)(\bcD^2\e^{-K/3})\notag\\
&&\qquad\quad~~-\frac18eM(\cD^2\e^{-K/3})+\I\frac14e(\bar\psi_a\bar\sigma^a)^{\alpha}(\cDla\e^{-K/3})
-\frac14e\big(\psi_{ab}\sigma^b\bar\psi^a+\I M\bar\psi_a\bar\sigma^a+\I\psi_a b^a\big)^\alpha(\cDla\e^{-K/3})\notag\\
&&\qquad\quad~~+\frac1{32}e \cD^2\bcD^2 (\S T)-\frac1{16}e \I(\bar\psi_a\bar\sigma^a)^\alpha\cDla\bcD^2(\S T)\notag\\
&&\qquad\quad~~+\frac18e \big(M^*+\bar\psi_a\bar\sigma^{ab}\bar\psi_b\big) \bcD^2 (\S T)+\frac1{24}eM \cD^2 (\S T)\Big]\Big|+{\rm h.c.} \notag\\
&&\qquad\quad~~+e\Big(-\frac12\cR-\frac13|M|^2+\frac13b^ab_a+\frac14\veps^{abcd}(\bar\psi_a\bar\sigma_b\psi_{cd}-\psi_a\sigma_b\bar\psi_{cd})\Big)\e^{-K(A,A^*)/3} \label{20} \\
&&\qquad\quad~~+eF\p W(A)+eF^*(\p W(A))^*-\frac12e\chi^2\p^2 W(A)-\frac12e\bar\chi^2(\p^2 W(A))^*\notag\\
&&\qquad\quad~~-\frac1{\sqrt2}e\I\chi\sigma^a\bar\psi_a\p W(A)-\frac1{\sqrt2}e\I\bar\chi\bar\sigma^a\psi_a(\p W(A))^* \nn \\
&&\qquad\quad~~ -e\left(M^*+\bar\psi_a\bar\sigma^{ab}\bar\psi_b\right)W(A)-e\left(M+\psi_a\sigma^{ab}\psi_b\right)W(A)^* \nn 
\end{eqnarray}
where ${\big|}$ specifies taking the lowest component of the superfield and 
\begin{equation}
{\psi_{mn}}^{\alpha}={\tilde{\cal{D}}}_{m}{\psi_{n}}^{\alpha}-{\tilde{\cal{D}}}_{n}{\psi_{m}}^{\alpha}, \quad   {\tilde{\cal{D}}}_{m}{\psi_{n}}^{\alpha}=\partial_{m}{\psi_{n}}^{\alpha}+{\psi_{n}}^{\beta}{\omega_{m\beta}}^{\alpha} \ .
\label{burt1}
\end{equation}
Since we are interested in the supergravity extension of the ghost condensate, we can, as in the globally supersymmetric case, set $W=0$. It then follows from their equations of motion that both
\begin{equation}
F= M=0 
\label{21}
\end{equation}
in our zero-fermion background.
This simplifies the Lagrangian \eqref{20}, which now becomes
\begin{eqnarray}
&&\mac{L}^{SUGRA} =\Big[-\frac3{32}e(\cD^2\bcD^2\e^{-K/3})+\I\frac3{16}e\bar\psi_{a\dot\alpha}\bar\sigma^{a\dot\alpha\alpha}(\cDla\bcD^2\e^{-K/3})-\frac38e\bar\psi_a\bar\sigma^{ab}\bar\psi_b(\bcD^2\e^{-K/3})\notag\\
&&\qquad\quad~~+\I\frac14e(\bar\psi_a\bar\sigma^a)^{\alpha}(\cDla\e^{-K/3})
-\frac14e\big(\psi_{ab}\sigma^b\bar\psi^a+\I\psi_a b^a\big)^\alpha(\cDla\e^{-K/3})\notag\\
&&\qquad\quad~~+\frac1{32}e \cD^2\bcD^2 (\S T)-\frac1{16}e \I(\bar\psi_a\bar\sigma^a)^\alpha\cDla\bcD^2(\S T)\notag\\
&&\qquad\quad~~+\frac18e\bar\psi_a\bar\sigma^{ab}\bar\psi_b \bcD^2 (\S T)\Big]\Big|+{\rm h.c.} \label{22} \\
&&\qquad\quad~~+e\Big(-\frac12\cR+\frac13b^ab_a+\frac14\veps^{abcd}(\bar\psi_a\bar\sigma_b\psi_{cd}-\psi_a\sigma_b\bar\psi_{cd})\Big)\e^{-K(A,A^*)/3} \ . \nn
\end{eqnarray}
Note that the auxiliary field $b_{a}=e_{a}^{m}b_{m}$ remains undetermined. To proceed, one must evaluate the lowest component of each superfield term.

Evaluating the first term in \eqref{19}, we find after integration by parts that
\begin{align}
 \frac1{e}\mac{L}^{SUGRA}_{K(\Phi,\Phi^{\dagger})}& = \frac1{e}\Big[\int \d^2\Theta 2\mac{E}\frac38(\bcD^2-8R)\e^{-K/3}\Big]+{\rm h.c.} \notag\\
&=\Big(-\frac12\cR+\frac13b^ab_a+\frac14\veps^{abcd}(\bar\psi_a\bar\sigma_b\psi_{cd}-\psi_a\sigma_b\bar\psi_{cd})\Big)\e^{-K(A,A^*)/3}\notag\\
&+3|\p A|^2(\e^{-K/3})_{,AA^*}+\I b^a\big(A_{,a}(\e^{-K/3})_{,A}-A^*{}_{,a}(\e^{-K/3})_{,A^*}\big)\notag\\
&-\I\frac1{\sqrt2}b^a\big(\psi_a\chi(\e^{-K/3})_{,A}-\bar\psi_a\bar\chi(\e^{-K/3})_{,A^*}\big)\notag\\
&-\sqrt2\chi\sigma^{mn}\psi_{mn}(\e^{-K/3})_{,A}-\sqrt2\bar\chi\bar\sigma^{mn}\bar\psi_{mn}(\e^{-K/3})_{,A^*}\label{23} \\
&-\I\frac32\psi_a\sigma^{ab}\sigma^c\bar\psi_bA_{,c}(\e^{-K/3})_{,A}-\I\frac32\bar\psi_a\bar\sigma^{ab}\bar\sigma^c\psi_bA^*{}_{,c}(\e^{-K/3})_{,A^*}\notag\\
&+\frac12\chi\sigma^a\bar\chi b_a(\e^{-K/3})_{,AA^*}+\I\frac32\big(\chi\sigma^a{e_a}^m\cD_m\bar\chi+\bar\chi\bar\sigma^a{e_a}^m\cD_m\chi\big)(\e^{-K/3})_{,AA^*}\notag\\
&+\frac32\sqrt2A^*{}_{,b}\psi_a\sigma^b\bar\sigma^a\chi(\e^{-K/3})_{,AA^*}+\frac32\sqrt2A_{,b}\bar\psi_a\bar\sigma^b\sigma^a\bar\chi(\e^{-K/3})_{,AA^*}\notag\\
&-\frac32(\p A)^2(\e^{-K/3})_{,AA}-\frac32(\p A^*)^2(\e^{-K/3})_{,A^*A^*}\notag\\
&+\I\frac32\chi\sigma^a\bar\chi\Big(A^*{}_{,a}(\e^{-K/3})_{,AA^*A^*}-A_{,a}(\e^{-K/3})_{,AAA^*}\Big) \nn
\end{align}
Note that this corresponds to the supergravitational $-X$ term in \eqref{12} if one chooses
\begin{equation}
K(\Phi,\Phi^{\dagger})=-\Phi\Phi^{\dagger} \ .
\label{24}
\end{equation}
The second term in \eqref{19} depends on the arbitrary hermitian function $T$. As a first step, let us choose $T=\tau/16$ where $\tau$ is a real constant.  For $\tau=1$, the second term in \eqref{19} corresponds to the supergravitational $X^{2}$ term in \eqref{12}. It is useful, however, to introduce $\tau$ as a ``marker'' indicating the component terms arising from this part of the Lagrangian. We will set $\tau=1$ at the end of the calculation. Evaluating this second term gives
\begin{align}
\frac1{e}\mac{L}^{SUGRA}_{\S,\tau}&= \frac{1}{e}\Big(-\frac{\tau}{2^{7}}\int \d^2\Theta 2\mac{E}(\bcD^2-8R)(\S)\Big) +{\rm h.c.} \notag\\
&=\Big(+\frac{\tau}{2^{9}}\cD^2\bcD^2 (\S)-\frac{\tau}{2^{8}}\I(\bar\psi_a\bar\sigma^a)^\alpha\cDla\bcD^2(\S)\notag\\
&+\frac{\tau}{2^{7}}\bar\psi_a\bar\sigma^{ab}\bar\psi_b\bcD^2 (\S)\Big)\Big|+{\rm h.c.}\notag\\
&={+\tau(\p A)^2(\p A^*)^2}-\frac12\sqrt2\tau\bar\psi_a\bar\sigma^a\sigma^c\bar\chi A^*{}_{,c}(\p A)^2-\frac12\sqrt2\tau\chi\sigma^c\bar\sigma^a\psi_aA_{,c}(\p A^*)^2\notag\\
&-\sqrt2\tau(\p A^*)^2A_{,m}\chi\psi^m-\sqrt2\tau(\p A)^2A^*{}_{,m}\bar\psi^m\bar\chi\notag \\
&{-\frac{\I}2\tau\chi\sigma^a\bar\chi A_{,a}{e_b}^m\cD_mA^*{}_{,b}}+\frac56\tau\chi\sigma^a\bar\chi A_{,a}A^*{}_{,b}b^b\label{XsquaredbeforeWeyl}\\
&{+\frac{\I}2\tau\chi\sigma^a\bar\chi A^*{}_{,a}{e_b}^m\cD_mA_{,b}}+\frac56\tau\chi\sigma^a\bar\chi A^*{}_{,a}A_{,b}b^b \notag \\
&{-\I\tau(\cD_m\chi)\sigma^b\bar\chi A^{,m}A^*{}_{,b}}+\sqrt2\tau\bar\psi_a\bar\sigma^c\sigma^b\bar\chi A^{,a}A^*{}_{,b}A_{,c}+\frac13\tau\bar\chi\bar\sigma^b\sigma_c\bar\sigma_a\chi b^c A^{,a}A^*{}_{,b}\notag \\
&{+\I\tau\chi\sigma^b(\cD_m\bar\chi)A^{*,m}A_{,b}}+\sqrt2\tau\chi\sigma^b\bar\sigma^c\psi_aA^{*,a}A_{,b}A^*{}_{,c}\notag\\
&{-\frac{\I}2\tau\chi\sigma^a\bar\sigma^b\sigma^m(\cD_m\bar\chi)A_{,a}A^*{}_{,b}}-\frac1{12}\tau\chi\sigma^a\bar\sigma^b\sigma^c\bar\chi b_cA_{,a}A^*{}_{,b}\notag\\
&{+\frac{\I}2\tau(\cD_m\chi) \sigma^m\bar\sigma^b\sigma^a\bar\chi A^*{}_{,a}A_{,b}}-\frac1{12}\tau\chi\sigma^c\bar\sigma^b\sigma^a\bar\chi b_cA^*{}_{,a}A_{,b} \ . \nn
\end{align}

The basic $N=1$ supergravity Lagrangian for the ghost condensate is obtained by adding \eqref{23} and \eqref{XsquaredbeforeWeyl}. Note that it contains the supergravity auxiliary field $b_{m},$ which can be eliminated from the Lagrangian using its equation of motion. This is found to be
\begin{align}
b_m=&-\frac32\I \left(A_{,m}(\e^{-K/3})_{,A}-A^*{}_{,m}(\e^{-K/3})_{,A^*}\right)\e^{K/3}-\frac34\chi\sigma_m\bar\chi(\e^{-K/3})_{,AA^*}\e^{K/3}\notag\\
&+\frac34\sqrt2\I\left(\psi_m\chi(\e^{-K/3})_{,A}-\bar\psi_m\bar\chi(\e^{-K/3})_{,A^*}\right)\e^{K/3}\notag\\
&-\frac54\tau\chi\sigma^a\bar\chi\big(A_{,a}A^*{}_{,m}+A^*{}_{,a}A_{,m}\big)\e^{K/3}\label{eom-b}\\
&+\frac12\tau\chi\sigma^a\bar\sigma_m\sigma^b\bar\chi A_{,a}A^*{}_{,b}\e^{K/3}\notag\\
&+\frac18\tau \big(\chi\sigma^a\bar\sigma^b\sigma_m\bar\chi+\chi\sigma_m\bar\sigma^a\sigma^b\bar\chi\big)A_{,a}A^*{}_{,b}\e^{K/3}\nn \ .
\end{align}
Plugging \eqref{eom-b} back into the sum of the Lagrangians, and keeping only the terms containing at most two fermions, this results in the expression
\begin{align}
\frac1{e}\mac{L}^{SUGRA}_{T={\tau}\slash16} &= \frac1{e}\int \d^2\Theta 2\mac{E}\Big[\frac38(\bcD^2-8R)\e^{-K/3}-\frac{\tau}{2^{7}}(\bcD^2-8R)(\S)\Big]+{\rm h.c.}\notag\\
&=\big(-\frac12\cR+\frac14\veps^{abcd}(\bar\psi_a\bar\sigma_b\psi_{cd}-\psi_a\sigma_b\bar\psi_{cd})\big)\e^{-K(A,A^*)/3}+3|\p A|^2(\e^{-K/3})_{,AA^*}\notag\\
&-\sqrt2\chi\sigma^{mn}\psi_{mn}(\e^{-K/3})_{,A}-\sqrt2\bar\chi\bar\sigma^{mn}\bar\psi_{mn}(\e^{-K/3})_{,A^*}\notag\\
&-\I\frac32\psi_a\sigma^{ab}\sigma^c\bar\psi_bA_{,c}(\e^{-K/3})_{,A}-\I\frac32\bar\psi_a\bar\sigma^{ab}\bar\sigma^c\psi_bA^*{}_{,c}(\e^{-K/3})_{,A^*}\notag\\
&+\I\frac32\big(\chi\sigma^a{e_a}^m\cD_m\bar\chi+\bar\chi\bar\sigma^a{e_a}^m\cD_m\chi\big)(\e^{-K/3})_{,AA^*}\notag\\
&+\frac32\sqrt2A^*{}_{,b}\psi_a\sigma^b\bar\sigma^a\chi(\e^{-K/3})_{,AA^*}+\frac32\sqrt2A_{,b}\bar\psi_a\bar\sigma^b\sigma^a\bar\chi(\e^{-K/3})_{,AA^*}\notag\\
&-\frac32(\p A)^2(\e^{-K/3})_{,AA}-\frac32(\p A^*)^2(\e^{-K/3})_{,A^*A^*}\notag\\
&+\I\frac32\chi\sigma^a\bar\chi\Big(A^*{}_{,a}(\e^{-K/3})_{,AA^*A^*}-A_{,a}(\e^{-K/3})_{,AAA^*}\Big)\notag\\
&{+\tau(\p A)^2(\p A^*)^2}
-\frac12\sqrt2\tau\bar\psi_a\bar\sigma^a\sigma^c\bar\chi A^*{}_{,c}(\p A)^2-\frac12\sqrt2\tau\chi\sigma^c\bar\sigma^a\psi_aA_{,c}(\p A^*)^2\notag\\
&-\sqrt2\tau(\p A^*)^2A_{,m}\chi\psi^m-\sqrt2\tau(\p A)^2A^*{}_{,m}\bar\psi^m\bar\chi \notag \\
&{-\frac{\I}2\tau\chi\sigma^a\bar\chi A_{,a}{e_b}^m\cD_mA^*{}_{,b}+\frac{\I}2\tau\chi\sigma^a\bar\chi A^*{}_{,a}{e_b}^m\cD_mA_{,b}}\label{25} \\
&{-\I\tau(\cD_m\chi)\sigma^b\bar\chi A^{,m}A^*{}_{,b}}+\sqrt2\tau\bar\psi_a\bar\sigma^c\sigma^b\bar\chi A^{,a}A^*{}_{,b}A_{,c}\notag\\
&{+\I\tau\chi\sigma^b(\cD_m\bar\chi)A^{*,m}A_{,b}}+\sqrt2\tau\chi\sigma^b\bar\sigma^c\psi_aA^{*,a}A_{,b}A^*{}_{,c}\notag \\
&{-\frac{\I}2\tau\chi\sigma^a\bar\sigma^b\sigma^m(\cD_m\bar\chi)A_{,a}A^*{}_{,b}+\frac{\I}2\tau(\cD_m\chi) \sigma^m\bar\sigma^b\sigma^a\bar\chi A^*{}_{,a}A_{,b}}\notag\\
&+\frac34\big((\p A)(\e^{-K/3})_{,A}-(\p A^*)(\e^{-K/3})_{,A^*}\big)^2\e^{K/3}\notag\\
&-\frac34\sqrt2\big(A^{,m}(\e^{-K/3})_{,A}-A^{*,m}(\e^{-K/3})_{,A^*}\big)\big(\psi_m\chi(\e^{-K/3})_{,A}-\bar\psi_m\bar\chi(\e^{-K/3})_{,A^*}\big)\e^{K/3}\notag\\
&-\frac34\I\chi\sigma^a\bar\chi\big(A_{,a}(\e^{-K/3})_{,A}-A^*{}_{,a}(\e^{-K/3})_{,A^*}\big)(\e^{-K/3})_{,AA^*}\e^{K/3}\notag\\
&-\frac74\I\tau\chi\sigma^a\bar\chi\big(A^*{}_{,a}(\p A)^2(\e^{-K/3})_{,A}-A_{,a}(\p A^*)^2(\e^{-K/3})_{,A^*}\big)\e^{K/3}\notag\\
&-\frac32\i\tau\chi\sigma^a\bar\chi\big(A_{,a}(\e^{-K/3})_{,A}-A^*{}_{,a}(\e^{-K/3})_{,A^*}\big)|\p A|^2\e^{K/3} \ . \nn
\end{align}
To go to Einstein frame and to render all fields canonically normalized, we now Weyl rescale as
\begin{align}
{e_n}^a&\weyl\e^{K/6}{e_n}^a \nn\\
\chi&\weyl\e^{-K/12}\chi \label{26}\\
\psi_m&\weyl\e^{K/12}\psi_m  \nn
\end{align}
and shift
\be
\psi_m\shift\psi_m+\I\frac{\sqrt2}6\sigma_m\bar\chi K_{,A^*} \ .
\label{27}
\ee
For the sum of terms not proportional to $\tau$, this results in 
\begin{align}
 \frac1{e}\mac{L}^{SUGRA}_{K(\Phi,\Phi^{\dagger}),{\rm Weyl}}= &\frac1{e}\Big[\int \d^2\Theta 2\mac{E}\frac38(\bcD^2-8R)\e^{-K/3}\Big]_{\rm Weyl}+{\rm h.c.}\notag\\
=&-\frac12\cR-K_{,AA^*}|\p A|^2\label{28}\\
&-\I K_{,AA^*}\bar\chi\bar\sigma^m\cD_m\chi+\veps^{klmn}\bar\psi_k\bar\sigma_l\tcD_m\psi_n\notag\\
&-\frac12\sqrt2K_{,AA^*}A^*{}_{,n}\chi\sigma^m\bar\sigma^n\psi_m-\frac12\sqrt2K_{,AA^*}A_{,n}\bar\chi\bar\sigma^m\sigma^n\bar\psi_m \ . \nn
\end{align}
See \cite{WB92} for details. After Weyl rescaling, the terms proportional to $\tau$ become
\begin{align}
\frac1{e}\mac{L}^{SUGRA}_{\S,\tau,{\rm Weyl}} &=\frac1{e}\Big[\int \d^2\Theta 2\mac{E}(-\frac{\tau}{2^{7}})(\bcD^2-8R)(\S)\Big]_{\rm Weyl}+{\rm h.c.}\notag\\
&={+\tau(\p A)^2(\p A^*)^2}-\frac12\sqrt2\tau\bar\psi_a\bar\sigma^a\sigma^c\bar\chi A^*{}_{,c}(\p A)^2-\frac12\sqrt2\tau\chi\sigma^c\bar\sigma^a\psi_aA_{,c}(\p A^*)^2\notag\\
&-\sqrt2\tau(\p A^*)^2A_{,m}\chi\psi^m-\sqrt2\tau(\p A)^2A^*{}_{,m}\bar\psi^m\bar\chi\notag\\
&{-\frac{\I}2\tau\chi\sigma^a\bar\chi A_{,a}e^{bm}(\cD_mA^*{}_{,b})+\frac{\I}2\tau\chi\sigma^a\bar\chi A^*{}_{,a}e^{bm}(\cD_mA_{,b})}\notag\\
&-\frac{\I}6\tau\chi\sigma^a\bar\chi A_{,a}A^*{}_{,b}K^{,b}+\frac{\I}6\tau\chi\sigma^a\bar\chi A^*{}_{,a}A_{,b}K^{,b} \nn\\
&{-\I\tau(\cD_m\chi)\sigma_n\bar\chi A^{,m}A^{*,n}}+\sqrt2\tau\bar\psi_a\bar\sigma^c\sigma^b\bar\chi A^{,a}A^*{}_{,b}A_{,c}\notag\\
&+\frac{\I}{12}\tau\chi\sigma^a\bar\chi A_{,b}A^*{}_{,a}K^{,b}+\frac{\I}6\tau\chi\sigma^{cb}\sigma^a\bar\chi A_{,c}A^*{}_{,a}K_{,b}\label{bla}\\
&{+\I\tau\chi\sigma^b(\cD_m\bar\chi)A^{*,m}A_{,b}}+\sqrt2\tau\chi\sigma^b\bar\sigma^c\psi_aA^{*,a}A_{,b}A^*{}_{,c}\notag\\
&-\frac{\I}{12}\tau\chi\sigma^a\bar\chi A^*{}_{,b}A_{,a}K^{,b}-\frac{\I}6\tau\chi\sigma^a\bar\sigma^{bc}\bar\chi A^*{}_{,c}A_{,a}K_{,b}\notag\\
&{-\frac{\I}2\tau\chi\sigma^p\bar\sigma^q\sigma^m(\cD_m\bar\chi)A_{,p}A^*{}_{,q}+\frac{\I}2\tau(\cD_m\chi)\sigma^m\bar\sigma^p\sigma^q\bar\chi A_{,p}A^*{}_{,q}}\notag\\
&+\frac{\I}6\tau\chi\sigma^c\bar\sigma^b\sigma^a\bar\chi K_{,a}A^*{}_{,b}A_{,c}-\frac{\I}6\tau\chi\sigma^a\bar\sigma^b\sigma^c\bar\chi K_{,a}A_{,b}A^*{}_{,c}\notag\\
&-\frac74\I\tau\chi\sigma^a\bar\chi\big(A^*{}_{,a}(\p A)^2(\e^{-K/3})_{,A}-A_{,a}(\p A^*)^2(\e^{-K/3})_{,A^*}\big)\e^{K/3}\notag\\
&-\frac32\I\tau\chi\sigma^a\bar\chi\big(A_{,a}(\e^{-K/3})_{,A}-A^*{}_{,a}(\e^{-K/3})_{,A^*}\big)|\p A|^2\e^{K/3}\nn \ .
\end{align}
To arrive at this result, we used 
\begin{align}
\cD_nA_{,b}&\weyl\e^{-K/6}\big(\cD_nA_{,b}-\frac16K_{,n}A_{,b}+\frac16{e_b}^lA^{,m}(K_{,m}g_{nl}-K_{,l}g_{nm})\big) \nn\\
\cD_n\chi^\alpha&\weyl\e^{-K/12}\big(\cD_n\chi^\alpha-\frac1{12}K_{,n}\chi^\alpha+\frac1{12}\chi^\beta(\sigma^{ml})_\beta{}^\alpha(K_{,m}g_{nl}-K_{,l}g_{nm})\big) \label{29}\\
\cD_n\bar\chi^{\dot\alpha}&\weyl\e^{-K/12}\big(\cD_n\bar\chi^{\dot\alpha}-\frac1{12}K_{,n}\bar\chi^{\dot\alpha}-\frac1{12}(\bar\sigma^{ml})^{\dot\alpha}{}_{\dot\beta}\bar\chi^{\dot\beta}(K_{,m}g_{nl}-K_{,l}g_{nm})\big) \ . \nn
\end{align}
This follows from the definitions 
\begin{equation}
\omega_{n\beta}{}^\alpha=\frac12(\sigma^{ml})_\beta{}^{\alpha}\omega_{nml}, \quad \cD_n\chi^\alpha=\p_n\chi^\alpha+\chi^\beta\omega_{n\beta}{}^\alpha
\label{burt2}
\end{equation}
 and the fact that under \eqref{26}
\begin{equation}
\omega_{nml}\weyl\e^{K/3}(\omega_{nml}+\frac16K_{,m}g_{nl}-\frac16K_{,l}g_{nm}) \ .
\label{30}
\end{equation}
The effect of the shift \eqref{27} on \eqref{bla} actually sums to zero.

\subsection*{B. The $N=1$ Supergravity Ghost Condensate}

The supergravity extension of the prototype scalar ghost condensate  $P(X)=-X+X^{2}$ is the sum of \eqref{28} and \eqref{bla}, where we take 
\begin{equation}
K(\Phi,\Phi^{\dagger})=-\Phi\Phi^{\dagger} , \qquad \tau=1 \ .
\label{31}
\end{equation}
That is,
\begin{equation}
\mac{L}^{SUGRA}_{T={1}\slash16,{\rm Weyl}} =\frac{1}{8}\Big[\int \d^2\Theta 2\mac{E}(\bcD^2-8R)\Big(3\e^{\Phi\Phi^{\dagger}/3}-\frac{1}{2^{4}}(\S)\Big)\Big]_{\rm Weyl}+{\rm h.c.}
\label{32}
\end{equation}
It follows from \eqref{28}, \eqref{bla} and \eqref{31} that the purely bosonic part of this Lagrangian is 
\begin{equation}
\frac{1}{e}\mac{L}^{SUGRA}_{T={1}\slash16,{\rm Weyl}} =-\frac12\cR+|\p A|^2{+(\p A)^2(\p A^*)^2}+ \dots
\label{33}
\end{equation}
Defining $A=\frac{1}{\sqrt{2}}(\phi+\i\xi)$, this becomes\footnote{Our conventions for gravity are adapted to those of Wess and Bagger \cite{WB92}: in terms of affine connections, the Riemann tensor is defined as $\cR_{mn}{}^p{}_s \equiv -\pt_m \Gamma^p_{ns}  + \pt_n \Gamma^p_{ms} - \Gamma^p_{mt}\Gamma^t_{ns} + \Gamma^p_{nt}\Gamma^t_{ms},$ and the Ricci tensor is given by $\cR_{mn} = \cR^p{}_{npm}.$ In terms of the spin connection, the Riemann tensor is $\cR_{mn}{}^{ab}\equiv \pt_m \omega_n{}^{ab} - \pt_n \omega_m{}^{ab} + \omega_{m}{}^{ac}\omega_{nc}{}^b - \omega_{n}{}^{ac}\omega_{mc}{}^b.$}
\begin{eqnarray}
&&\frac{1}{e}\mac{L}^{SUGRA}_{T={1}\slash16,{\rm Weyl}} =-\frac12\cR+\frac{1}{2}(\pt\phi)^2 + \frac{1}{4}(\pt\phi)^4 \label{34} \\
&&\quad\qquad\qquad\qquad+\frac{1}{2} (\pt\xi)^2 + \frac{1}{4}(\pt\xi)^4 - \frac{1}{2}(\pt\phi)^2(\pt\xi)^2 + (\pt\phi\cdot\pt\xi)^2 +\dots\nn
\end{eqnarray}
The remaining terms in the Lagrangian are at least quadratic in the fermions $\chi$, $\psi_{m}$. 
The Einstein and gravitino equations can be solved in a flat FRW spacetime $ds^{2}=-dt^{2}+a(t)^{2}\delta_{ij}dx^{i}dx^{j}$ with a vanishing gravitino $\psi_{m}=0$.
The $\phi$, $\xi$ and $\chi$ equations of motion continue to admit a ghost condensate vacuum of the form
 \begin{equation}
\phi=ct~, \quad \xi=0~, \quad \chi=0
 \label{36}
 \end{equation}
where, to be consistent with the coupling to dynamical $a(t)$, one must set $c=1$. 
The scale factor is that of a de Sitter spacetime, which--in its flat slicing--is given by
\be
a(t) = e^{\pm \frac{1}{\sqrt{12}}t} \ .
\ee
The choice of the $\pm$ sign corresponds to an expanding or contracting space respectively; in this paper, we focus on the expanding branch. 
To assess the stability of the supergravity ghost condensate, one can expand in small fluctuations around this background. Considering scalar fluctuations 
\be \phi = t + \delta\phi(t,\vec{x})\,, \qquad \xi = \delta\xi(t,\vec{x})
\label{37}
\ee
only, the result to quadratic order is
\bea
\frac{1}{e}\mac{L}^{SUGRA}_{T={1}\slash16,{\rm Weyl}} &=& (\dot{\delta\phi})^2 + \, 0 \cdot \delta\phi^{,i}\delta\phi_{,i} \nn \\ &+& \, 0 \cdot (\dot{\delta\xi})^2 + \delta\xi^{,i}\delta\xi_{,i} \label{39}  \ . 
\eea
As in the globally supersymmetric case, both lines illustrate important issues to be addressed in supergravity ghost condensation--that is, the $\delta\phi$ spatial gradient instability and the unacceptable $\delta\xi$ kinetic terms respectively. We will present the solution to both of these problems later in the paper. Now, however, we turn to a discussion of the fermions in the background of the ghost condensate.

\subsection*{C. The Fermion Lagrangian and the Super-Higgs Effect}

For a discussion of the fermions in the ghost condensate vacuum, the relevant part of Lagrangian \eqref{32} is\footnote{To find the ghost condensate background, it is consistent to set the auxiliary fields $M=F=0$ since $M$ and $F$ are sourced only by terms of quadratic and higher order in fermions. However, one might wonder whether it is necessary to include the $M$- and $F$-terms in the calculation of quadratic fermionic fluctuations around this background. Luckily, we do not have to do this. In the absence of a superpotential, all terms arising from the substitution of $M$ and $F$ into the action are fourth-order and higher in fermions and, hence, do not contribute to the present calculation. This follows from the results of Appendix B and from the analysis of the equation of motion for $F$ detailed in \cite{KLO12}.}
\begin{align}
\frac{1}{e}\mac{L}^{SUGRA}_{T={1}\slash16,{\rm Weyl}} & =\dots+ \frac12\veps^{klmn}\Big(\bar\psi_k\bar\sigma_l\tcD_m\psi_n-\psi_k\sigma_l\tcD_m\bar\psi_n\Big)\notag\\
&+\frac{\I}2\big(\chi\sigma^m\cD_m\bar\chi+\bar\chi\bar\sigma^m\cD_m\chi\big)\left(1+\frac12(\p\phi)^2\right)\label{40}\\
& +\frac{\I}2\phi^{,m}\phi_{,n}\big(\bar\chi\bar\sigma^n(\cD_m\chi)+\chi\sigma^n(\cD_m\bar\chi)\big)\notag\\
&+\frac12\left(\chi\sigma^m\bar\sigma^n\psi^p+\bar\chi\bar\sigma^m\sigma^n\bar\psi^p\right)\left(g_{mp}\phi_{,n}+\frac12g_{mn}\phi_{,p}(\p\phi)^2-\frac12g_{np}\phi_{,m}(\p\phi)^2\right)+\dots \nn
\end{align}
where $g_{mn}$ is the FRW metric.
For the time-dependent vev $\phi=t$, $(\p\phi)^2 = -1$. Hence, the first and second/third lines correspond to unmixed $\psi_{m}$ and $\chi$ kinetic energies respectively. However, the ghost condensate
induces a mass mixing term between $\chi$ and $\psi_m.$ Using $\sigma^m\bar\sigma^n = - g^{mn} + 2 \sigma^{mn}$, the mass term can be rewritten as 
\be
\frac14 \phi_{,m}(\chi\psi^m + \bar\chi\bar\psi^m) - \frac12 \phi_{,m}(\chi\sigma^{mn}\psi_n + \bar\chi\bar\sigma^{mn}\bar\psi_n)
\label{41}
\ee
or, more simply, 
\be
-\frac14 \phi_{,m}(\chi\sigma^{m}\bar\sigma^n\psi_n + \bar\chi\bar\sigma^{m}\sigma^n\bar\psi_n).
\label{42}
\ee

Let us try to eliminate this mass mixing by redefining the gravitino. As we will discuss below, the supersymmetry transformations suggest the field redefinition
\begin{equation}
\psi_{m\alpha}=\tilde\psi_{m\alpha}-\frac{2\I}{(\pt\phi)^2}\cD_m(\phi_{,n}\sigma^n_{\alpha\dot\alpha}\bar\chi^{\dot\alpha}) \ .
\label{43}
\end{equation}
Using the fact that the second partial derivative on $\phi$ vanishes, the $\psi_{m}$ kinetic term transforms into
\begin{align}
&\frac12\veps^{klmn}\Big(\bar\psi_k\bar\sigma_l\tcD_m\psi_n-\psi_k\sigma_l\tcD_m\bar\psi_n\Big)=\frac12\veps^{klmn}\Big(\tilde{\bar\psi}_k\bar\sigma_l\tcD_m\tilde\psi_n-\tilde\psi_k\sigma_l\tcD_m\tilde{\bar\psi}_n\Big)\notag\\
&\quad\qquad\qquad\qquad+2\I\veps^{klmn}\Big(\tbp_k\bar\sigma_l\cD_m\cD_n(\phi_{,p}\sigma^p\bar\chi)-\tp_k\sigma_l\cD_m\cD_n(\phi_{,p}\bar\sigma^p\chi)\Big)\label{44}\\
&+2\veps^{klmn}\Big(\cD_k(\chi\sigma^p\phi_{,p})\bar\sigma_l\cD_m\cD_n(\phi_{,q}\sigma^q\bar\chi)-\cD_k(\bar\chi\bar\sigma^p\phi_{,p})\sigma_l\cD_m\cD_n(\phi_{,q}\bar\sigma^q\chi)\Big) \ .\nn
\end{align}
Furthermore, employing  the relation
\be
(\cD_m \cD_n - \cD_n \cD_m) \chi=-\frac{\cR}{12}\sigma_{mn}\chi,
\label{45}
\ee
which is valid for maximally symmetric spacetimes, and the fact that $\cR=-1$ for our de Sitter background, the $\psi_{m}$ kinetic term becomes
\begin{align}
&\frac12\veps^{klmn}\Big(\bar\psi_k\bar\sigma_l\tcD_m\psi_n-\psi_k\sigma_l\tcD_m\bar\psi_n\Big)=\frac12\veps^{klmn}\Big(\tilde{\bar\psi}_k\bar\sigma_l\tcD_m\tilde\psi_n-\tilde\psi_k\sigma_l\tcD_m\tilde{\bar\psi}_n\Big)\notag\\
&~~\quad\quad\qquad\qquad\qquad\qquad\qquad\qquad+\frac14\phi_{,p}\Big(\tbp_k\bar\sigma^k\sigma^p\bar\chi+\tp_k\sigma^k\bar\sigma^p\chi\Big)\label{46}\\
&\quad +\frac{\I}4\Big(-(\p\phi)^2\cD_k\chi\sigma^k\bar\chi+2\phi_{,p}\phi^{,k}\cD_k\chi\sigma^p\bar\chi-(\p\phi)^2\cD_k\bar\chi\bar\sigma^k\chi+2\phi_{,p}\phi^{,k}\cD_k\bar\chi\bar\sigma^p\chi\Big) \ . \nn
\end{align}
Since we are only working to quadratic order in fermions, the second term on the right-hand side can be written as
\begin{equation}
\frac14\phi_{,p}\Big(\tbp_k\bar\sigma^k\sigma^p\bar\chi+\tp_k\sigma^k\bar\sigma^p\chi\Big)=+\frac14 \phi_{,m}(\chi\sigma^{m}\bar\sigma^n\psi_n + \bar\chi\bar\sigma^{m}\sigma^n\bar\psi_n)+\dots \ ,
\label{47}
\end{equation}
where we have anti-commuted the fermions, used the definition of ${\bar{\sigma}}^{m}$ and relabeled indices. Note that this term {\it exactly cancels} the $\chi$, $\psi_{m}$ mass mixing term \eqref{42}. Furthermore, the remaining terms in \eqref{46} do {\it not} introduce mixing of the ${\tilde{\psi}}_{m}$, $\chi$ kinetic energies. It follows that in the ghost condensate vacuum, using  $(\p\phi)^2=-1$ and the redefined gravitino ${\tilde{\psi}}_{m}$, the quadratic fermion terms in \eqref{40} reduce to 
\begin{align}
\frac{1}{e}\mac{L}^{SUGRA}_{T={1}\slash16,{\rm Weyl}} &=\dots+\frac12\veps^{klmn}\Big(\tbp_k\bar\sigma_l\tcD_m\tp_n-\tp_k\sigma_l\tcD_m\tbp_n\Big)\notag\\
&+\frac{\I}2\big(\chi\sigma^m\cD_m\bar\chi+\bar\chi\bar\sigma^m\cD_m\chi\big)\label{48}\\
&+\I \phi^{,m}\phi_{,n}\left(\bar\chi\bar\sigma^n(\cD_m\chi)+\chi\sigma^n(\cD_m\bar\chi)\right) +\dots\notag
\end{align}
This Lagrangian describes a) a {\it massless} gravitino ${\tilde{\psi}}_{m}$ with Lorentz covariant kinetic energy and b) a {\it massless} fermion $\chi$ with kinetic terms whose Lorentz invariance is broken in the ghost condensate background.
We note that after the field redefinition of the gravitino, the kinetic terms for $\chi$ now appear with an additional overall multiplicative factor of $2.$ 

Given this result, one can analyze the super-Higgs effect within the context of the supergravity ghost condensate. We know from the discussion in Subsection III C that the ghost condensate spontaneously breaks global $N=1$ supersymmetry. What happens when this is generalized to supergravity? We showed in \cite{KLO11} and Appendix A that the variations of the fermions $\chi$ and $\psi_{m}$ under local supersymmetry--after  Weyl rescaling and using the solutions for the supergravity auxiliary fields $M$ and $b_{m}$ appropriate to a bosonic background--are given by
\bea
\delta \chi &=& \I \sqrt{2} \sigma^m \bar{\zeta} \pt_m A + \sqrt{2} e^{K/6} \zeta F\ , \label{EqVarChi} \\ 
\delta \psi_m &=& 2 \big(\cD_m + \frac14 (K_{,A}\pt_m A-K_{,A^{*}}\pt_m A^{*})\big)  \zeta+ \I e^{K/2}W\sigma_m \bar{\zeta} \ , \label{49}
\eea
for arbitrary K\"ahler potential $K$, superpotential $W$ and chiral auxiliary field $F$. Since we are interested in supersymmetry breaking in the vacuum, we have ignored all terms proportional to the component fermions on the right-hand side of the variations. In {\it pure two-derivative} chiral theories coupled to supergravity--that is, not in the ghost condensate case--spontaneous breaking of supersymmetry is achieved as follows.  One chooses a non-vanishing $W$ for which 1) the potential energy is minimized by having the scalar $A$ be a constant, and 2) when evaluated at this 
minimum $F=-K^{,AA^*}e^{K/3}(D_AW)^* \neq 0$, where $D_A W$ is the K\"ahler covariant derivative of $W$. The non-vanishing $F$-term in \eqref{EqVarChi} then renders the $\chi$ transformation inhomogeneous, spontaneously breaking supersymmetry, while the transformation of a redefined gravitino ${\tilde{\psi}}_{m}$ vanishes. Therefore, $\chi$ is the massless Goldstone fermion while ${\tilde{\psi}}_{m}$ is the physical gravitino. Generically, $W\neq0$ in the vacuum giving the gravitino a non-vanishing mass 
\begin{equation}
m_{3/2}=e^{K/2}|W| \ .
\label{50}
\end{equation}
As first discussed in \cite{Cremmer:1978hn}, in the process the Goldstone fermion $\chi$ gets ``eaten'' by the now massive gravitino.
This is the super-Higgs effect. Note, however, that if $W=0$ in the vacuum--but 
with $DW\neq 0$--the gravitino mass vanishes even though supersymmetry is spontaneously broken. Although this is generically not the case, it is possible to find theories where this does occur.

Let us now return to the supergravity ghost condensate vacuum. In this case we choose the holomorphic function $W=0$, from which it follows that $F=0$. However, $A$ now develops a non-zero, linearly time-dependent vev $\langle A \rangle = \langle \phi \rangle/\sqrt{2}=ct/\sqrt{2}$, where we restore the dimension-two constant $c$. The $\chi$ transformation in \eqref{EqVarChi} then becomes
\bea
\delta \chi &=& \i \sqrt{2} \sigma^m \bar\zeta \pt_m A =\i \s^0 \bar\zeta c \ .
\label{51}
\eea
As previously, the fermion transforms inhomogeneously and, hence, supersymmetry is spontaneously broken. 
For the ghost condensate, however, the inhomogeneous term arises from the linear time-dependent vev of $\phi$ rather than from the $F$-term. Now consider the gravitino transformation \eqref{49}. Recalling that we choose $K=-\Phi\Phi^{\dagger}$ in the ghost condensate, and using $W=0$ and $\langle A \rangle=ct/\sqrt{2}$, it follows from \eqref{49} that
\begin{equation}
\delta\psi_{m}= 2 \big(\cD_m + \frac14 (K_{,A}\pt_m A-K_{,A^{*}}\pt_m A^{*})\big) \zeta=2{\cal{D}}_{m}\zeta \ .
\label{52}
\end{equation}
Note that, in addition to the term proportional to $W$ vanishing, the factor $K_{,A}\pt_m A-K_{,A^{*}}\pt_m A^{*}$ in the first term is also zero in this vacuum. Be this as it may, the de Sitter spacetime covariant derivative ${\cal{D}}_{m}\zeta_\a=\partial_{m}\zeta_\a -\frac12 \omega_{mpl} (\sigma^{pl})_\a{}^\beta \zeta_\beta$ does not vanish, as
\begin{equation}
\omega_{i0j}=g_{ij}H ,
\label{WB1}
\end{equation}
and, hence, $\psi_{m}$ transforms inhomogeneously. However, in analogy with the ordinary two-derivative case, let us redefine the gravitino as in \eqref{43}. It is straightforward to shown that in the ghost condensate background
\begin{equation}
\delta {\tilde{\psi}}_{m} = 0 \ .
\label{WB2}
\end{equation}
This then identifies $\chi$ as the massless Goldstone fermion and ${\tilde{\psi}}_{m}$ as the physical gravitino. The generic expression for the gravitino mass was given by \eqref{50}. In the ghost condensate, however, $W=0$ and, hence, 
\begin{equation}
m_{3/2}=0 \ .
\label{53}
\end{equation}
That is, the breaking of local supersymmetry  via a ghost condensate is analogous to two-derivative supergravity theories with a superpotential for which $DW \neq 0$ but $W=0$ in the vacuum.
This result for the supergravity ghost condensate is completely consistent with--and gives a physical explanation for--the above calculation of the quadratic fermion Lagrangian \eqref{48}. There we found, after appropriate redefinition of the gravitino, that the mixed $\chi$, ${\tilde{\psi}}_{m}$
mass terms exactly cancelled and that there were no diagonal $\chi\chi$ or ${\tilde{\psi}}{\tilde{\psi}}$ masses--exactly as expected from the variations \eqref{51},\eqref{WB2} and \eqref{53}.

\subsection*{D. Scalar Field Stability Analysis}

Recall from \eqref{39} that, when expanded around the ghost condensate vacuum, the quadratic $\delta\phi$ part of the Lagrangian is 
\begin{equation}
\frac{1}{e}\mac{L}^{SUGRA}_{T={1}\slash16,{\rm Weyl}} = (\dot{\delta\phi})^2 + \, 0 \cdot \delta\phi^{,i}\delta\phi_{,i} +\dots \label{57}  \ . 
\end{equation}
This is analogous to the globally supersymmetric case discussed in Subsection III B and, for the same reasons as discussed there, $\phi$ will develop a small, negative spatial gradient term in the NEC violating region where $P_{,X}<0$. This problem was overcome in the global supersymmetry case by adding the term \eqref{gradientstabilizer2} to ${\cal{L}}^{SUSY}$. It is straightforward to generalize this to the supergravity case with the addition of the term
\begin{equation}
-\frac18\int \d^2\Theta2\cE(\bcD^2-8R)(\S T_{\phi}) +{\rm h.c.}
\label{58}
\end{equation}
where 
\begin{equation}
T_{\phi} =\frac{\kappa}{2^{9}}\Big(\{\cDua,\bcDla\}\{\cDla,\bcDua\}(\Phi+\Phid)\Big)^2 
\label{59}
\end{equation}
and where $\kappa$ is a real number. Note that in Subsection III B we (somewhat arbitrarily) set the parameter $\kappa=-1/4$. This reflected the fact that, in the globally supersymmetric case, the exact value of this parameter is irrelevant to the discussion. However, as we will see, this is not the case when coupled to supergravity.
We calculate \eqref{58},\eqref{59} in terms of component fields for $F=M=0$ and to quadratic order in fermions $\chi$ and $\psi_{m}$ in Appendix B. It suffices here to present only those terms required to analyze the existence and stability of the ghost condensate. These are
\begin{align}
-\frac{1}{8e}\Big[\int \d^2\Theta&2\cE(\bcD^2-8R)\S T_{\rm \phi}\Big]_{\rm Weyl} +{\rm h.c.} \notag\\
=& \kappa (\Box\phi)^2 \left( (\partial\phi)^4 + (\partial\xi)^4 - 2 (\partial\phi)^2(\partial\xi)^2 + 4(\partial\phi \cdot \partial\xi)^2\right) \ . 
\label{60}
\end{align}
The remaining terms are at least quadratic in the fermions $\chi$ and $\psi_{m}$. 
When this is added to the Lagrangian \eqref{34}, the equations of motion for the component fields are modified. We restrict our attention to gravity and the scalar $\phi$, since these are the only non-vanishing fields in the ghost condensate background. The relevant part of the Lagrangian is
\be
\frac{1}{\sqrt{-g}} \mac{L} = -\frac{\cR}{2} + \frac12 (\partial\phi)^2 + \frac14 (\p\phi)^4 + \kappa (\p\phi)^4 (\Box\phi)^2.
\ee
The associated equations of motion are
\bea
0 &=& -\Box\phi \big(1+(\p\phi)^2\big) - 2 \phi^{;mn}\phi_{,m}\phi_{,n}-8\kappa \phi^{;mn}\phi_{,m}\phi_{,n}(\Box\phi)^2 - 4 \kappa (\p\phi)^2 (\Box\phi)^3 \nn \\ && - 8 \kappa (\p\phi)^2 \Box\phi \, {\phi_{;n}}^{nm}\phi_{,m} + 16 \kappa \phi^{;pn}\phi_{;mn}\phi_{,p}\phi^{,m}\Box\phi + 8 \kappa (\p\phi)^2 \Box\phi \, \phi^{;mn}\phi_{;mn} \nn \\ && + 8 \kappa (\p\phi)^2 \Box\phi  \, \phi^{,n} \phi_{;nm}{}^m + 8 \kappa (\p\phi)^2 \phi^{,n} \phi_{;nm} \phi_{;p}{}^{pm} \nn \\ && + 8 \kappa (\p\phi)^2 \phi^{;pm} \phi_{,p}\phi^{;n}{}_{nm} + 2 \kappa (\p\phi)^4 \phi_{;n}{}^n{}_m{}^m \label{XX} \ , \\
G_{mn} &=& \phi_{,m} \phi_{,n} \big(1+ (\p\phi)^2 + 4 \kappa (\p\phi)^2 (\Box\phi)^2\big) \nn \\ && - \frac12 g_{mn} (\p\phi)^2 \big(1+ \frac12 (\p\phi)^2 - 2 \kappa (\p\phi)^2 (\Box\phi)^2 -16\kappa \Box \phi \, \phi^{;rs} \phi_{,r} \phi_{,s} - 4 \kappa (\p\phi)^2 \phi_{;s}{}^{sr} \phi_{,r}\big) \nn \\ && - 8 \kappa (\p\phi)^2 \Box\phi \, \phi^{,r}(\phi_{;rm}\phi_{,n} + \phi_{;rn} \phi_{,m}) - 2 \kappa (\p\phi)^4 (\phi^{;r}{}_{rm}\phi_{,n} + \phi^{;r}{}_{rn} \phi_{,m}) \label{YY} 
\eea
where $G_{mn}$ is the Einstein tensor\footnote{To derive the Einstein equations, the identity  \bea
\frac{\delta}{\delta g^{mn}} \int \sqrt{-g} f \Box \phi &=& \int \sqrt{-g} \big(-\frac{1}{2}g_{mn} f \Box \phi + \frac{\delta f}{\delta g^{mn}} \Box \phi+ f \phi_{;mn} \nn \\ && \qquad \qquad-\frac{1}{2}\nabla_m (f\phi_{,n}) - \frac{1}{2} \nabla_n (f \phi_{,m}) + \frac{1}{2} g_{mn} \nabla^p (f\phi_{,p})\big) 
\eea
is useful--where $f$ is a scalar function of the fields.
The first term on the right-hand side arises from varying $\sqrt{-g}$, while the second line comes from varying the metric inside of the connection in the $\Box\phi$ term.}.
We are interested in the question of whether these equations of motion still admit a ghost condensate/de Sitter solution. Therefore, we look for a solution where $\dot\phi$ is constant and the metric is a de Sitter space with constant Hubble parameter $H.$ With this Ansatz, the equations of motion greatly simplify to
\bea
&&0 = 1 - \dot\phi^2 - 9 \kappa \dot\phi^6 + 6 \kappa \dot\phi^8, \label{ZZ}\\ 
&&12 H^2 = 3 \dot\phi^2 - 2 \dot\phi^4. \label{X1}
\eea
The first equation is quartic in $\dot\phi^2,$ where the solution of interest is the one that reduces to $\dot\phi^2 = 1$ as $\kappa \rightarrow 0.$ This solution then allows one to calculate the Hubble rate using the second equation. For small $\kappa$--which, from an effective field theory point of view, is the case of real interest--a perturbative solution is easy to derive. It is given by
\bea
\langle\dot\phi\rangle^2 &=& 1 - 3 \kappa + {\cal{O}}(\kappa^2), \label{GCdS1} \\
\langle H \rangle^2 &=& \frac{1}{12} + \frac14 \kappa + {\cal{O}}(\kappa^2). \label{GCdS2}
\eea
Thus, the effect of adding the stabilizing term for $\phi$ is to shift the parameters of the ghost condensate/de Sitter solution without altering its qualitative features\footnote{One might ask what the solution becomes for large $\kappa$. By inspection, we see that in this regime the solution is approximately $\dot\phi^2 \approx 3/2$ with $H^2$ very small. Thus, for large $\kappa,$ one obtains a kind of ghost condensate in Minkowski spacetime. However, one should refrain from taking the $(\Box\phi)^2$ term too seriously when $\kappa$ is large--since it leads to fourth-order equations of motion. Hence, it only makes sense from an effective field theory point of view, in which case its coefficient must be small for consistency.}. We now explicitly demonstrate the stability of $\phi.$ Expanding about this new vacuum using \eqref{37}, the $\phi$ part of the component field Lagrangian becomes
\be
{\cal L}^{\rm SUGRA} = \frac12 (3 \langle\dot\phi\rangle^2 -1) (\dot{\delta\phi})^2 +  \frac{1}{2a^2} (1- \langle\dot\phi\rangle^2) \delta\phi^{,i}\delta\phi_{,i} + \kappa (\Box \delta\phi)^2 + \dots \ .
\label{fluctuations3}
\ee
For small $\kappa$, this leads to the dispersion relation
\be
\omega^2 \approx -\kappa\big(\frac32k^2 + k^4\big). \label{disp1}
\ee
Thus, to tame instabilities, one must require 1) $\kappa<0$ and 2) that $|\kappa|$ be sufficiently large.
For a discussion of the allowed phenomenological range of $\kappa,$ see \cite{BKO07}. Happily, the required values of $|\kappa|$ are still sufficiently small to allow the above perturbative expansion. To apply ghost condensate theory to models of a bouncing universe, one introduces a potential which causes $\langle \dot\phi \rangle^2$ to be slightly lowered. This has the consequence that the NEC is then violated. In this case, it may happen that the $k^2$ term in the dispersion relation \eqref{disp1} switches sign. This signals a gradient instability at long wavelengths and, correspondingly, the bounce must occur on a fast time-scale. However, at short wavelength (large $k$) one can see that the introduction of the $(\Box\phi)^2$ term indeed stabilizes the ghost condensate.

We now turn our attention to the second scalar, $\xi.$ The second line 
\begin{equation}
\frac{1}{e}\mac{L}^{SUGRA}_{T={1}\slash16,{\rm Weyl}} = \dots+ \, 0 \cdot (\dot{\delta\xi})^2 + \delta\xi^{,i}\delta\xi_{,i}+ \dots \label{60A}  \ . 
\end{equation}
in \eqref{39} indicates that, when expanded around the ghost condensate,  the time derivative term in the $\delta \xi$ kinetic energy 
vanishes, while the spatial gradient term has the wrong sign. This result is analogous to the globally supersymmetric case discussed in Subsection III B, and was cured by adding the supersymmetric higher-derivative terms \eqref{walk1} to ${\cal{L}}^{SUSY}$. It is straightforward to generalize this to the supergravity case by adding
\begin{equation}
-\frac18\int \d^2\Theta2\cE(\bcD^2-8R)(\S T_{\xi}) +{\rm h.c.},
\label{61}
\end{equation}
where
%
\bea
T_{\xi}&=+(1+6\kappa) 2^{-5}\{\cDua,\bcDla\}(\Phi-\Phid)\{\cDla,\bcDua\}(\Phid-\Phi)\notag\\
&~~-(1+9\kappa)2^{-10}\Big(\{\cDua,\bcDla\}(\Phi+\Phid)\{\cDla,\bcDua\}(\Phi-\Phid)\Big)^2,
 \label{62}
\eea
to ${\cal{L}}^{SUGRA}$. Note that we have multiplied each of the two terms by an independent real coefficient - the reason for our particular choice of coefficients will become clear momentarily. One can calculate \eqref{61},\eqref{62} in terms of component fields for $F=M=0$ and to quadratic order in fermions $\chi$ and $\psi_{m}$ . It suffices here to present only those terms required to analyze the existence and stability of the ghost condensate. These are
\begin{align}
-\frac{1}{8e}\Big[\int \d^2\Theta&2\cE(\bcD^2-8R)\S T_{\xi}\Big]_{\rm Weyl} +{\rm h.c.} \notag\\
&=-2(1+6\kappa)\,(\pt\phi)^4(\pt\xi)^2 - (1+9\kappa)\,(\pt\phi)^4(\pt\phi\cdot\pt\xi)^2 \ . 
\label{63}
\end{align}
The remaining terms are at least quadratic in the fermions $\chi$ and $\psi_{m}$. When these are added to the Lagrangian, the modified equations of motion continue to admit the same ghost condensate/de Sitter vacuum as the one derived above in Eqs. (\ref{GCdS1}) and (\ref{GCdS2}). Expanding around this vacuum using \eqref{37}, and taking into account the new stabilizing term in \eqref{58}, the fluctuation Lagrangian for $\xi$ becomes
\begin{eqnarray}
&&\frac{1}{e}{\cal L}^{\rm SUGRA} = \dots +\Big(-\frac12 + \frac12 \langle\dot\phi\rangle^2 + 2 (1+6\kappa) \langle\dot\phi\rangle^4 - (1+9\kappa) \langle\dot\phi\rangle^6 +2\kappa \langle \Box\phi\rangle^2 \langle \dot\phi \rangle^2  \Big) (\dot{\delta \xi})^2 \nn \\ && \qquad\qquad\qquad\quad + \Big(\frac12 + \frac12 \langle\dot\phi\rangle^2 - 2 (1+6\kappa) \langle\dot\phi\rangle^4  +2\kappa \langle \Box\phi\rangle^2 \langle \dot\phi \rangle^2 \Big) \delta\xi^{,i}\delta\xi_{,i}+ \dots \nn \\
&&\qquad\qquad~ =\dots  +\Big(1 + {\cal{O}}(\kappa^{2})\Big)\Big( (\dot{\delta \xi})^2 
- \delta\xi^{,i}\delta\xi_{,i}\Big)+ \dots 
\label{65}
\end{eqnarray}
Thus the scalar $\xi$ is rendered completely stable by the addition of these terms. Moreover, our choice of coefficients in \eqref{62} implies that for small $|\kappa|$ the fluctuations are canonical. 

\subsection*{E. The Modified Fermion Lagrangian and Super-Higgs Effect}

Having resolved the $\delta \phi$ spatial gradient and $\delta \xi$ wrong sign kinetic problems in the supergravity context, one must re-examine the question of the fermion Lagrangian and the super-Higgs effect in the presence of the additional terms \eqref{58},\eqref{59} and \eqref{61},\eqref{62}. In principle, this is a difficult calculation, requiring the evaluation of all terms quadratic in the fermions $\chi$ and $\psi_{m}$. As can be seen, for example, by examining the $T_{\phi}$ Lagrangian in Appendix B, although some fermion terms vanish in the ghost condensate vacuum, some, both kinetic and mass terms, are non-zero. Evaluating each of these, inserting them into the complete supergravity Lagrangian and then diagonalizing all fermion kinetic energy and mass terms is a lengthy undertaking. Happily, to understand the essential physics, it is unnecessary to carry this out.

Recall from the discussion in Subsection IV C that one can decide the fermion masses by analyzing the behavior of their transformations under local supersymmetry. In \eqref{51} and \eqref{52} we presented the supersymmetry transformations in the ghost condensate situation where $W=F=M=b_{m}=0$. Since \eqref{36} continues to be valid, and since (in a bosonic background) the $b_{m}$ equation of motion is unchanged by the higher-derivative terms \eqref{58},\eqref{59} and \eqref{61},\eqref{62},
it follows that the $\chi$ and $\psi_{m}$ variations remain
\bea
\delta \chi &=& \i \sqrt{2} \sigma^m \bar\zeta \pt_m A =\i \s^0 \bar\zeta c 
\label{66}
\eea
and
\begin{equation}
\delta \psi_{m}=2{\cal{D}}_{m}\zeta \ .
\label{69}
\end{equation}
respectively. 
As previously, it is straightforward to define a new physical gravitino ${\tilde{\psi}}_{m}$ which transforms homogeneously. The required definition is given by Eq. (\ref{43}) but where now $\dot\phi$ and the connection $\omega_m$ are evaluated in the shifted vacuum. Since the fermion transformation \eqref{66} is inhomogeneous, supersymmetry is spontaneously broken with a massless Goldstone fermion $\chi$. Furthermore, since $W=0$ in the ghost condensate vacuum, the mass of the physical gravitino ${\tilde{\psi}}_{m}$ is 
\begin{equation}
m_{3/2}=0 \ .
\label{WB3}
\end{equation}
We can conclude from these arguments that, even in the presence of the additional terms, the quadratic fermion Lagrangian will describe a {\it massless} Goldstone fermion $\chi$ and a {\it massless} gravitino ${\tilde{\psi}}_{m}$ with diagonal kinetic energies.

\begin{acknowledgements}

M.K. and J.L.L. gratefully acknowledge the support of the European Research Council via the Starting Grant numbered 256994. B.A.O. is supported in part by the DOE under contract No. DE-AC02-76-ER-03071 and the NSF under grant No. 1001296.

\end{acknowledgements}

\appendix

\section{The Weyl Rescaled Fermion Supersymmetry Transformations}

Prior to Weyl rescaling, the fermion supersymmetry transformations--see equations 18.23 and 19.14 in Wess and Bagger \cite{WB92}--are given by
\bea
\delta \chi &=& \I \sqrt{2} \sigma^m \bar{\zeta} \partial_m A + \sqrt{2} \zeta F, \label{B1}\\
\delta \psi_m &=& 2\cD_m \zeta - \I {e_m}^a \left(\frac13 M \sigma_a\bar{\zeta} + b_a \zeta + \frac13 b^c \zeta \sigma_c\bar\sigma_a \right), \label{B2}
\eea
where we have dropped all component fermions on the right-hand side of the variations since these vanish in the vacua of interest and $\zeta$ is the supersymmetry parameter.  Note that our parameter is minus the one in equations 18.23 and 19.14 of Wess and Bagger--a convention adopted later in their book. Weyl rescaling is performed via
\bea
e_{n}^{a} &\weyl& e^{K/6} e_{n}^{a} \ , \nn \\
\chi &\weyl& e^{-K/12} \chi \ , \label{B3} \\ \psi_m &\weyl& e^{K/12} \psi_m  \nn
\eea
and
\begin{equation}
\zeta \weyl e^{K/12} \zeta \ . 
\label{B4}
\end{equation}
Then the Weyl rescaled variations are
\bea
&&e^{-K/12}\delta \chi_{\text{\tiny WEYL}} = \I \sqrt{2} \sigma^a {e_a}^m e^{-K/6} \bar{\zeta} e^{K/12} \partial_m A + \sqrt{2} \zeta e^{K/12} F, \label{B5}\\
&&e^{K/12} \delta \psi_{m \text{\tiny WEYL}} = 2 e^{K/12} \Big(\cD_m \zeta + \frac{1}{12}K_{,m} \zeta -\frac16 (\zeta \sigma^{nl}) K_{,n} g_{ml}\Big) \label{B6}  \\ &&\qquad\qquad~\qquad - \I {e_m}^a e^{K/6} \left(\frac13 M \sigma_a\bar{\zeta}e^{K/12} + b_n {e_a}^n e^{-K/6} \zeta e^{K/12} + \frac13 {e_c}^n b_n e^{-K/6} \zeta \sigma^c\bar\sigma_a e^{K/12} \right) \nn
\eea
It is important to note that there are additional terms that arise from Weyl rescaling the covariant derivative $\cD_m \zeta^\a = \p_n\zeta^\alpha+\zeta^\beta\omega_{n\beta}{}^\alpha$ with $\omega_{n\beta}{}^\alpha=\frac12(\sigma^{ml})_\beta{}^{\alpha}\omega_{nml}$ using
\be
\omega_{nml}\weyl\e^{K/3}(\omega_{nml}+\frac16K_{,m}g_{nl}-\frac16K_{,l}g_{nm}).
\label{B7}
\ee
As discussed previously, the gravitino must also be shifted as
\bea 
\psi_m \shift \psi_m + \I \frac{\sqrt{2}}{6}K_{,A^{*}}\sigma_m \bar{\chi}
\label{B8}
\eea
in order for the fermionic kinetic terms to be in canonical form. For the supersymmetry transformation of $\psi_{m}$, this means that
\be
\delta \psi_{m \text{\tiny WEYL}} \rightarrow \delta \psi_{m\text{\tiny WEYL+SHIFT}} + \I \frac{\sqrt{2}}{6}K_{,A^{*}} \sigma_{m} \delta \bar\chi_{\text{\tiny WEYL}} \ .
\label{B9}
\ee
Therefore
\begin{align}
\delta \psi_{m \text{\tiny WEYL+SHIFT}}^\a &= \delta \psi_{m \text{\tiny WEYL}}^\a - \I \frac{\sqrt{2}}{6}K_{,A^{*}} \epsilon^{\a\gamma}\sigma_{m\gamma\dot\beta} \delta \bar\chi_{\text{\tiny WEYL}}^{\dot\beta} \notag \\
&= 2 \Big(\cD_m \zeta^\a + \frac{1}{12}K_{,m} \zeta^\a -\frac16 \zeta^\beta {(\sigma^{nl})_{\beta}}^{\a} K_{,n} g_{ml}\Big) \notag \\ 
& - \I  \left(\frac13 M e^{K/6}\epsilon^{\a\gamma}\sigma_{m\gamma\dot\beta}\bar{\zeta}^{\dot\beta} + b_m \zeta^\a + \frac13 b^c \zeta^\gamma \sigma_{c\gamma \dot\beta}\bar\sigma_m^{\dot\beta\a} \right) \notag \\ 
& -\I \frac{\sqrt{2}}{6}K_{,A^{*}} \epsilon^{\a\gamma}\sigma_{m\gamma\dot\beta} \Big(-\I \sqrt{2} \zeta^\delta \epsilon^{\dot\beta\dot\gamma}\sigma^n_{\delta\dot\gamma} \partial_n A^* + \sqrt{2} \bar\zeta^{\dot\beta} e^{K/6} F^*\Big)\notag \\ 
&= 2 \Big(\cD_m \zeta^\a - \frac{1}{12} \zeta^\beta \sigma^n_{\beta \dot\beta} \bar\sigma_m^{\dot\beta \a}K_{,n}\Big) \label{B10} \\ 
& - \I \left(\frac13 M e^{K/6}\epsilon^{\a\gamma}\sigma_{m\gamma\dot\beta}\bar{\zeta}^{\dot\beta} + b_m \zeta^\a + \frac13 b^c \zeta^\gamma \sigma_{c\gamma \dot\beta}\bar\sigma_m^{\dot\beta\a} \right) \notag \\ 
& +\frac13  \zeta^\beta \sigma^n_{\beta \dot\beta} \bar\sigma_m^{\dot\beta \a}K_{,A^*}\p_n A^* - \frac13 \I e^{K/6} \epsilon^{\a\gamma}\sigma_{m\gamma\dot\beta}\bar{\zeta}^{\dot\beta} K_{,A^*} F^* \notag \\
&= 2 \cD_m \zeta^\a -\frac16 \zeta^\beta \sigma^n_{\beta \dot\beta} \bar\sigma_m^{\dot\beta \a} (K_{,A}\p_n A - K_{,A^*} \p_n A^*) \notag \\ 
& - \frac13 \I e^{K/6} \epsilon^{\a\gamma}\sigma_{m\gamma\dot\beta}\bar{\zeta}^{\dot\beta} (M+K_{,A^*} F^*)  - \I  \left(b_m \zeta^\a + \frac13 b^n \zeta^\gamma \sigma_{n\gamma \dot\beta}\bar\sigma_m^{\dot\beta\a} \right). \nn
\end{align}

In the case of pure two-derivative chiral supergravity coupled to a superpotential, the solutions for $F$, $M$ and $b_m$ are given by 
\bea
&& F = - K^{,AA^*} e^{K/3} (D_A W)^* \label{B11}\\ && M+ K_{,A^*} F^* = N = -3 e^{K/3} W \label{B12}\\ && b_m = \frac{\I}{2}(K_{,A} \p_m A - K_{,A^*} \p_m A^*). \label{B13}
\eea
Plugging these solutions into \eqref{B5} and \eqref{B10}, we obtain 
\bea
\delta \chi_{\text{\tiny WEYL}} &=& \I \sqrt{2} \sigma^m \bar{\zeta} \partial_m A - \sqrt{2} K^{,AA^*} e^{K/2} (D_A W)^* \zeta \ , \label{EqVarChi2} \\ 
\delta \psi_{m \text{\tiny WEYL+SHIFT}} &=& 2 \Big(\cD_m + \frac14 (K_{,A}\partial_m A-K_{,A^{*}}\partial_m A^{*})\Big)  \zeta+ \I e^{K/2}W\sigma_m \bar{\zeta} \ .
\label{B14}
\eea
These reproduce the $\chi$ and $\psi_m$  supersymmetry variations given in equations 23.5 and 23.6 of \cite{WB92}. 

For the higher-derivative supergravity Lagrangians coupled to a superpotential introduced in \cite{KLO12}--and used to discuss the ghost-condensate vacuum in this paper--the solutions of the $M$ and $b_{m}$ equations of motion, when all component fermions are set to zero, continue to be given by \eqref{B12} and \eqref{B13}. This was proven in \cite{KLO12} for any higher-derivative addition to the Lagrangian of the form $\S T$, where $T$ is an arbitrary hermitian function of $\Phi,\Phi^{\dagger}$ with any number of their spacetime derivatives. For example, note that in the $T=\tau/16$ case discussed in Subsection IV A of this paper, the solution for the $b_{m}$ equation of motion is given in \eqref{eom-b}. When the component fermions are set to zero, this becomes 
\begin{equation}
b_m=-\frac32\I \left(A_{,m}(\e^{-K/3})_{,A}-A^*{}_{,m}(\e^{-K/3})_{,A^*}\right)\e^{K/3}= \frac{\I}{2}(K_{,A} \p_m A - K_{,A^*} \p_m A^*)
\label{cup1}
\end{equation}
which is identical to \eqref{B13}.
However, as discussed in detail in \cite{KLO12,KLO12b}, the equation of motion for the auxiliary field $F$ is now generically cubic and is no longer solved by \eqref{B11}. Putting \eqref{B12}, \eqref{B13} into \eqref{B5} and \eqref{B10}, but for an arbitrary solution $F$, the fermion variations become
\bea
\delta \chi_{\text{\tiny WEYL}} &=& \I \sqrt{2} \sigma^m \bar{\zeta} \pt_m A + \sqrt{2} e^{K/6} \zeta F\ , \label{EqVarChiA} \\ 
\delta \psi_{m \text{\tiny WEYL+SHIFT}} &=& 2 \big(\cD_m + \frac14 (K_{,A}\pt_m A-K_{,A^{*}}\pt_m A^{*})\big)  \zeta+ \I e^{K/2}W\sigma_m \bar{\zeta}  \label{B14A}
\eea
for any K\"ahler potential $K$ and superpotential $W$. These are the transformations used in \eqref{EqVarChi} and \eqref{49} in the text to analyze supersymmetry breaking and the fermion masses in the supergravitational ghost-condensate theory.

\section{Component Expansions} 

In this Appendix, we provide details about the component expansions of the higher-derivative superfield expressions that we employ in this paper. For completeness and potential future use, we will at first keep the terms that involve the auxiliary fields $M$ and $F.$ Note that we work only to quadratic order in fermions throughout.
The component expansion of a general higher-derivative term in our formalism is given by
\begin{align}
-&\frac1{8e}\int \d^2\Theta2\cE(\bcD^2-8R)(\S\phone)+h.c.\notag\\
=&+16\big\{(\p A)^2(\p A^*)^2-2|\p A|^2|F|^2+|F|^4\big\}\lc\phone\notag \\
&-4\sqrt2\big\{(\p A)^2A^*{}_{,b}\bar\psi_a\bar\sigma^a\sigma^b\bar\chi+(\p A^*)^2A_{,b}\psi_a\sigma^a\bar\sigma^b\chi\big\}\lc\phone+\I4\sqrt2|F|^2\big\{F^*\bar\psi_a\bar\sigma^a\chi+F\psi_a\sigma^a\bar\chi\big\}\lc\phone\notag\\
&+4\sqrt2|F|^2\big\{A_{,b}\bar\psi_a\bar\sigma^a\sigma^b\bar\chi+A^*{}_{,b}\psi_a\sigma^a\bar\sigma^b\chi\big\}\lc\phone+\I4\sqrt2A_{,b}A^*{}_{,c}\big\{F^*\bar\psi_a\bar\sigma^a\sigma^c\bar\sigma^b\chi+F\psi_a\sigma^a\bar\sigma^b\sigma^c\bar\chi\big\}\lc\phone\notag\\
&+\I8b_d\big\{F^*A^{*,d}\chi^2-FA^{,d}\bar\chi^2\big\}\lc\phone-8\big\{F_{,d}A^{,d}\bar\chi^2+F^*{}_{,d}A^{*,d}\chi^2\big\}\lc\phone\notag \\
&+8\big\{F^*\chi^2{e_a}^m\cD_mA^*{}_{,a}+F\bar\chi^2{e_a}^m\cD_mA_{,a}\big\}\lc\phone-16\sqrt2\big\{(\p A)^2A^{*,a}\bar\chi\bar\psi_a+(\p A^*)^2A^{,a}\psi_a\chi\big\}\lc\phone\notag\\
& -8\chi\sigma^a\bar\chi b_a|F|^2+2^3\I\chi\sigma^a\bar\chi\big\{FF^*{}_{,a}-F^*F_{,a}\big\}\lc\phone+\frac{40}3e\chi\sigma^b\bar\chi b^a\big\{A_{,a}A^*{}_{,b}+A_{,b}A^*{}_{,a}\big\}\lc\phone\notag \\
& +\I8\chi\sigma^a\bar\chi\big\{A^*{}_{,a}{e_b}^m\cD_mA{}_{,b}-A_{,a}{e_b}^m\cD_mA^*{}_{,b}\big\}\lc\phone-\I16A_{,a}A^*{}_{,b}\big\{(\hat{D}^a\chi)\sigma^b\bar\chi+(\hat{D}^a\bar\chi)\bar\sigma^b\chi\big\}\lc\phone\notag\\
&+\frac{16}3\chi\sigma^a\bar\sigma^c\sigma^b\bar\chi A_{,a}A^*{}_{,b}b_c\lc\phone-\I24|F|^2\big\{\chi\sigma^c(\hat{D}_c\bar\chi)+\bar\chi\bar\sigma^c(\hat{D}_c\chi)\big\}\lc\phone\notag \\
& +32\big\{F^*A^*{}_{,a}\chi\sigma^{ab}(\hat{D}_b\chi)+FA_{,a}\bar\chi\bar\sigma^{ab}(\hat{D}_b\bar\chi)\big\}\lc\phone-\I8A_{,a}A^*{}_{,b}\big\{\chi\sigma^a\bar\sigma^b\sigma^c(\hat{D}_c\bar\chi)+\bar\chi\bar\sigma^b\sigma^a\bar\sigma^c(\hat{D}_c\chi)\big\}\lc\phone\notag\\
&-\I\frac83\veps^{abcd}\chi\sigma_d\bar\chi A_{,a}A^*{}_{,b}b_c\lc\phone+\frac83\chi\sigma^a\bar\chi b_a|\p A|^2\lc\phone-2^4\sqrt2|F|^2\big\{A^{,a}\bar\chi\bar\psi_a+A^{*,a}\psi_a\chi\big\}{}\phone\big\vert\notag\\
&+\I8\sqrt2\big\{(\p A)^2A^*{}_{,b}(\bar\chi\bar\sigma^b)^\a{}\cDla\phone\big\vert+(\p A^*)^2A_{,a}(\chi\sigma^a)_{\dot\a}{}\bcDua\phone\big\vert\big\}\notag\\
&-8\sqrt2A_{,a}A^*{}_{,b}\big\{F^*(\chi\sigma^a\bar\sigma^b)^\a{}\cDla\phone\big\vert+F(\bar\chi\bar\sigma^b\sigma^a)_{\dot\a}{}\bcDua\phone\big\vert\big\}\notag\\
&-\I8\sqrt2|F|^2\big\{A_{,a}(\bar\chi\bar\sigma^a)^\a{}\cDla\phone\big\vert+A^*{}_{,a}(\chi\sigma^a)_{\dot\alpha}{}\bcDua\phone\big\vert\big\}-8\sqrt2|F|^2\big\{F^*\chi^\alpha{}\cDla\phone\big\vert+F\bar\chi_{\dot\alpha}{}\bcDua\phone\big\vert\big\}\notag\\
&+2(\p A)^2\bar\chi^2{}\cDDa\phone\big\vert+2(\p A^*)^2\chi^2{}\bcDDa\phone\big\vert-2(F^*)^2\chi^2{}\cDDa\phone\big\vert-2F^2\bar\chi^2{}\bcDDa\phone\big\vert\notag\\
&+\I4\chi\sigma^a\bar\chi\big\{F^*A_{,a}{}\cDDa\phone\big\vert-FA^*{}_{,a}{}\bcDDa\phone\big\vert\big\}\notag\\
&-\I2\big\{F^*A^*{}_{,a}\chi^2-FA_{,a}\bar\chi^2\big\}\big\{{}\sigma^a_{\alpha\dot\alpha}\cDua\bcDua\phone\big\vert+{}\sigma^a_{\alpha\dot\alpha}\cDua\bcDua\phone\big\vert^\dagger\big\}\notag\\
&-4A_{,a}A^*{}_{,b}(\chi\sigma^a)_{\dot\alpha}\big\{{}\cDua\bcDua\phone\big\vert+{}\cDua\bcDua\phone\big\vert^\dagger\big\}(\sigma^b\bar\chi)_\a\notag\\
&-4|F|^2\big\{\chi^\alpha{}\cDla\bcDua\phone\big\vert\bar\chi_{\dot\alpha}+\bar\chi_{\dot\alpha}{}\cDla\bcDua\phone\big\vert^\dagger\chi^\alpha\big\}\label{C1}
\end{align}
With $M$ and $F$ set to zero, this expression reduces to
\begin{align}
-&\frac1{8e}\int\d^2\Theta2\cE(\bcD^2-8R)(\S\phone)_{M=F=0}+h.c.\notag\\
=&+16(\p A)^2(\p A^*)^2\lc\phone+\I8\chi\sigma^a\bar\chi\big\{A^*{}_{,a}{e_b}^m\cD_mA{}_{,b}-A_{,a}{e_b}^m\cD_mA^*{}_{,b}\big\}\lc\phone\notag\\
&-4\sqrt2\big\{(\p A)^2A^*{}_{,b}\bar\psi_a\bar\sigma^a\sigma^b\bar\chi+(\p A^*)^2A_{,b}\psi_a\sigma^a\bar\sigma^b\chi\big\}\lc\phone+\frac83\chi\sigma^a\bar\chi b_a|\p A|^2\lc\phone\notag\\
&-16\sqrt2\big\{(\p A)^2A^{*,a}\bar\chi\bar\psi_a+(\p A^*)^2A^{,a}\psi_a\chi\big\}\lc\phone-\I16A_{,a}A^*{}_{,b}\big\{(\hat{D}^a\chi)\sigma^b\bar\chi+(\hat{D}^a\bar\chi)\bar\sigma^b\chi\big\}\lc\phone\notag\\
&+\frac{16}3\chi\sigma^a\bar\sigma^c\sigma^b\bar\chi A_{,a}A^*{}_{,b}b_c\lc\phone+\frac{40}3e\chi\sigma^b\bar\chi b^a\big\{A_{,a}A^*{}_{,b}+A_{,b}A^*{}_{,a}\big\}\lc\phone\notag \\
&-\I8A_{,a}A^*{}_{,b}\big\{\chi\sigma^a\bar\sigma^b\sigma^c(\hat{D}_c\bar\chi)+\bar\chi\bar\sigma^b\sigma^a\bar\sigma^c(\hat{D}_c\chi)\big\}\lc\phone-\I\frac83\veps^{abcd}\chi\sigma_d\bar\chi A_{,a}A^*{}_{,b}b_c\lc\phone\notag\\
&+\I8\sqrt2\big\{(\p A)^2A^*{}_{,b}(\bar\chi\bar\sigma^b)^\a{}\cDla\phone\big\vert+(\p A^*)^2A_{,a}(\chi\sigma^a)_{\dot\a}{}\bcDua\phone\big\vert\big\}\notag\\
&+2(\p A)^2\bar\chi^2{}\cDDa\phone\big\vert+2(\p A^*)^2\chi^2{}\bcDDa\phone\big\vert\notag\\
&-4A_{,a}A^*{}_{,b}(\chi\sigma^a)_{\dot\alpha}\big\{{}\cDua\bcDua\phone\big\vert+{}\cDua\bcDua\phone\big\vert^\dagger\big\}(\sigma^b\bar\chi)_\a
\end{align}

The stabilizing terms that we require in order for the scalar field fluctuations to be well-behaved correspond to the choice
\begin{align}
\phone=&-2^{-11}\Big[\{\cDua,\bcDla\}\{\cDla,\bcDua\}(\Phi+\Phid)\Big]^2\notag\\
&+2^{-5}\{\cDua,\bcDla\}(\Phi-\Phid)\{\cDla,\bcDua\}(\Phid-\Phi)\notag\\
&-2^{-10}\Big[\{\cDua,\bcDla\}(\Phi+\Phid)\{\cDla,\bcDua\}(\Phi-\Phid)\Big]^2\\
=&-2^{-5}\Big[\cD^a\cD_a(\Phi+\Phid)\Big]^2+2^{-2}\cD^a(\Phi-\Phid)\cD_a(\Phid-\Phi)\notag\\
&-2^{-4}\Big[\cD^a(\Phi+\Phid)\cD_a(\Phi-\Phid)\Big]^2\\
=&-2^{-5}\Big[\cD^a\cD_a\Phi+\cD^a\cD_a\Phid\Big]^2-2^{-2}\Big[\cD^a\Phi\cD_a\Phi+\cD^a\Phid\cD_a\Phid-2\cD^a\Phi\cD_a\Phid\Big]\notag\\
&-2^{-4}\Big[\cD^a\Phi\cD_a\Phi-\cD^a\Phid\cD_a\Phid\Big]^2.
\end{align}
We will split this up according and first consider( cf.~\eqref{59}),
\begin{align}
\phone_\phi\equiv&\frac{\kappa}{2^{9}}\Big(\{\cDua,\bcDla\}\{\cDla,\bcDua\}(\Phi+\Phid)\Big)^2 =\frac{\kappa}{8}\Big[\cD^a\cD_a(\Phi+\Phid)\Big]^2
\end{align}
Since we restrict to terms with at most two fermions overall, we see from \eqref{C1} that we need to evaluate the expressions $\lc{\cD^a\cD_a\Phi}_\tf$, $\lc{\cDla\cD^a\cD_a\Phi}_\of$, $\lc{\cDDa\cD^a\cD_a\Phi}_\zf$ to the order in fermions indicated by the subscript ({\it e.g.} ``2f'' standing for ``two fermions''). We obtain
\begin{align}
\lc{\cD^a\cD_a\Phi}_\tf=&{e^{am}\cD_mA_{,a}-\frac{\I}{12}\sqrt2M\chi\sigma^a\bar\psi_a-\frac12\sqrt2(\hD_a\chi)\psi^a}\notag\\
&{-\frac{\I}6\sqrt2\psi^a\chi b_a+\frac{\I}{24}\sqrt2\psi_a\sigma^a\bar\sigma^c\chi b_c}\\
\lc{\cDla\cD^a\cD_a\Phi}_\of=&{\frac19\sqrt2|M|^2\chi_\alpha+\frac{\I}3\sqrt2b_b\big\{\delta_\alpha{}^\beta\eta^{ab}-(\sigma^a\bar\sigma^b)_\alpha{}^\beta\big\}(\hD_a\chi_\beta)}\notag\\
&{+\frac{\I}6\sqrt2\big(e^{am}\cD_mb_b\big)\big\{\delta_\alpha{}^\beta\eta^{ab}-(\sigma^a\bar\sigma^b)_\alpha{}^\beta\big\}\chi_\beta-\frac{\I}6Fb_b\big\{\delta_\alpha{}^\beta\eta^{ab}-(\sigma^b\bar\sigma^a)_\alpha{}^\beta\big\}\psi_{a\beta}}\notag\\
&{-\frac16A_{,d}\sigma^d_{\alpha\dot\alpha}b_b\big\{\delta^{\dot\alpha}{}_{\dot\beta}\eta^{ab}-(\bar\sigma^b\sigma^a)^{\dot\alpha}{}_{\dot\beta}\big\}\bar\psi^{\dot\beta}_a-\frac1{36}\sqrt2\chi_\alpha b^ab_a+\sqrt2e^{am}\cD_m\hD_a\chi_\alpha}\notag\\
&{+\frac16M^*A_{,b}(\sigma^b\bar\sigma^a\psi_a)_\alpha-\psi_{a\alpha}F^{,a}-\frac{\I}6MF(\sigma_a\bar\psi^a)_\alpha-\I\big(e^{am}\cD_mA_{,b}\big)(\sigma^b\bar\psi^a)_\alpha}\\
\lc{\cDla\cD^a\cD_a\Phid}_\of=&{-\frac{\sqrt2}6M^*b_a(\sigma^a\bar\chi)_\alpha-\frac{\I}6\sqrt2M^*(\sigma^a\hD_a\bar\chi)_\alpha-\frac{\I}6\sqrt2M^*{}_{,a}(\sigma^a\bar\chi)_\alpha}\\
\lc{\cDDa\cD^a\cD_a\Phi}_\zf=&{\frac{16}9Fb^ab_a-\frac89F|M|^2+\frac{16}9\I A_{,a}b^aM^*-\frac{16}3\I b^aF_{,a}-\frac43M^*e^{am}\cD_mA_{,a}}\notag\\
&{-\frac83\I Fe^{am}\cD_mb_a-\frac23A^{,m}M^*{}_{,m}-4e^{am}\cD_mF_{,a}}\\
\lc{\cDDa\cD^a\cD_a\Phid}_\zf=&{\frac89F(M^*)^2+\frac49\I A^*{}_{,a}b^aM^*+\frac43M^*e^{am}\cD_mA^*{}_{,a}+\frac23A^{*,m}M^*{}_{,m}}\\
\lc{\cDla\bcDla\cD^a\cD_a\Phi}_\zf=&{\Big[-\frac29MFb_a-\I\frac29|M|^2A_{,a}+\I\frac23MF_{,a}+\frac{\I}3M_{,a}F\Big]\sigma^a_{\alpha\dot\alpha}}\\
\lc{\cDla\bcDla\cD^a\cD_a\Phid}_\zf=&{\Big[-\frac29M^*F^*b_a-\I\frac49|M|^2A^*{}_{,a}-\I\frac23M^*F^*{}_{,a}-\frac{\I}3M^*{}_{,a}F^*}\notag\\
&{-\I\frac49b^db_dA^*{}_{,a}+\I\frac49A^*{}_{,c}b^cb_a-2\I e^{bm}\cD_m({e_b}^n\cD_nA^*{}_{,a})}\notag\\
&{+\I\frac43\veps^{a'bcd}\eta_{aa'}b_c{e_d}^m\cD_mA^*{}_{,b}+\I\frac23\veps^{a'bcd}\eta_{aa'}A^*{}_{,b}{e_d}^m\cD_mb_c\Big]\sigma^a_{\alpha\dot\alpha}}
\end{align}
Then
\begin{align}
\lc{\phone_\phi}_\tf=&\frac{\kappa}{8}\Big\{-\Big(e^{am}\cD_m(A_{,a}+A^*{}_{,a})\Big)^2\notag\\
&-8\kappa e^{bm}\cD_mA_{,b}\lc{\cD^a\cD_a\Phi}_\tf+2e^{bm}\cD_mA^*{}_{,b}\lc{\cD^a\cD_a\Phid}_\tf\notag\\
&-8\kappa e^{bm}\cD_mA_{,b}\lc{\cD^a\cD_a\Phid}_\tf+2e^{bm}\cD_mA^*{}_{,b}\lc{\cD^a\cD_a\Phi}_\tf\Big\}\\
\lc{\cDla\phone_\phi}_\of=&\frac{\kappa}{4} e^{bm}\cD_m(A_{,b}+A^*{}_{,b})\Big\{\lc{\cDla\cD^a\cD_a\Phi}_\of+\lc{\cDla\cD^a\cD_a\Phid}_\of\Big]\\
\lc{\cDDa\phone_\phi}_\zf=&\frac{\kappa}{4} e^{bm}\cD_m(A_{,b}+A^*{}_{,b})\Big\{\lc{\cDDa\cD^a\cD_a\Phi}_\zf+\lc{\cDDa\cD^a\cD_a\Phid}_\zf\Big]\\
\lc{\cDla\bcDla\phone_\phi}_\zf=& \frac{\kappa}{4} e^{bm}\cD_m(A_{,b}+A^*{}_{,b})\Big\{\lc{\cDla\bcDla\cD^a\cD_a\Phi}_\zf+\lc{\cDla\bcDla\cD^a\cD_a\Phid}_\zf\Big\}
\end{align}
i.e.
\begin{align}
\lc{\phone_\phi}_\tf=&\frac{\kappa}{4} \Big(e^{bm}\cD_m(A_{,b}+A^*{}_{,b})\Big)\Big\{\frac12e^{am}\cD_m(A_{,a}+A^*{}_{,a})\notag\\
&-\frac{\I}{12}\sqrt2M\chi\sigma^a\bar\psi_a-\frac12\sqrt2(\hD_a\chi)\psi^a\notag\\
&{-\frac{\I}6\sqrt2\psi^a\chi b_a+\frac{\I}{24}\sqrt2\psi_a\sigma^a\bar\sigma^c\chi b_c}\notag\\
&{-\frac{\I}{12}\sqrt2M^*\bar\chi\bar\sigma^a\psi_a-\frac12\sqrt2\bar\psi^a(\hD_a\bar\chi)}\notag\\
&+\frac{\I}6\sqrt2\bar\chi\bar\psi^a b_a-\frac{\I}{24}\sqrt2\bar\psi_a\bar\sigma^a\sigma^c\bar\chi b_c\Big\}\\
\lc{\cDla\phone_\phi}_\text{1f}=& \frac{\kappa}{4} e^{cm}\cD_m(A_{,c}+A^*{}_{,c})\Big\{\frac19\sqrt2|M|^2\chi_\alpha+\frac{\I}3\sqrt2b_b\big\{\delta_\alpha{}^\beta\eta^{ab}-(\sigma^a\bar\sigma^b)_\alpha{}^\beta\big\}(\hD_a\chi_\beta)\notag\\
&{+\frac{\I}6\sqrt2\big(e^{am}\cD_mb_b\big)\big\{\delta_\alpha{}^\beta\eta^{ab}-(\sigma^a\bar\sigma^b)_\alpha{}^\beta\big\}\chi_\beta-\frac{\I}6Fb_b\big\{\delta_\alpha{}^\beta\eta^{ab}-(\sigma^b\bar\sigma^a)_\alpha{}^\beta\big\}\psi_{a\beta}}\notag\\
&{-\frac16A_{,d}\sigma^d_{\alpha\dot\alpha}b_b\big\{\delta^{\dot\alpha}{}_{\dot\beta}\eta^{ab}-(\bar\sigma^b\sigma^a)^{\dot\alpha}{}_{\dot\beta}\big\}\bar\psi^{\dot\beta}_a-\frac1{36}\sqrt2\chi_\alpha b^ab_a+\sqrt2e^{am}\cD_m\hD_a\chi_\alpha}\notag\\
&{+\frac16M^*A_{,b}(\sigma^b\bar\sigma^a\psi_a)_\alpha-\psi_{a\alpha}F^{,a}-\frac{\I}6MF(\sigma_a\bar\psi^a)_\alpha-\I\big(e^{am}\cD_mA_{,b}\big)(\sigma^b\bar\psi^a)_\alpha}\notag\\
&{-\frac{\sqrt2}6M^*b_a(\sigma^a\bar\chi)_\alpha-\frac{\I}6\sqrt2M^*(\sigma^a\hD_a\bar\chi)_\alpha-\frac{\I}6\sqrt2M^*{}_{,a}(\sigma^a\bar\chi)_\alpha}\Big\}\\
\lc{\cDDa\phone_\phi}_\text{0f}=& \frac{\kappa}{4} e^{cm}\cD_m(A_{,c}+A^*{}_{,c})\Big\{{\frac{16}9Fb^ab_a-\frac89F|M|^2+\frac{16}9\I A_{,a}b^aM^*}\notag\\
&-\frac{16}3\I b^aF_{,a}-\frac43M^*e^{am}\cD_mA_{,a}\notag\\
&{-\frac83\I Fe^{am}\cD_mb_a-\frac23A^{,m}M^*{}_{,m}-4e^{am}\cD_mF_{,a}}\notag\\
&{+\frac89F(M^*)^2+\frac49\I A^*{}_{,a}b^aM^*+\frac43M^*e^{am}\cD_mA^*{}_{,a}+\frac23A^{*,m}M^*{}_{,m}}\Big\}\\
\lc{\cDla\bcDla\phone_\phi}_\text{0f}=& \frac{\kappa}{4} e^{em}\cD_m(A_{,e}+A^*{}_{,e})\Big\{-\frac29MFb_a-\I\frac29|M|^2A_{,a}+\I\frac23MF_{,a}+\frac{\I}3M_{,a}F\notag\\
&-\frac29M^*F^*b_a-\I\frac49|M|^2A^*{}_{,a}-\I\frac23M^*F^*{}_{,a}-\frac{\I}3M^*{}_{,a}F^*\notag\\
&-\I\frac49b^db_dA^*{}_{,a}+\I\frac49A^*{}_{,c}b^cb_a-2\I e^{bm}\cD_m({e_b}^n\cD_nA^*{}_{,a})\notag\\
&+\I\frac43\veps^{a'bcd}\eta_{aa'}b_c{e_d}^m\cD_mA^*{}_{,b}+\I\frac23\veps^{a'bcd}\eta_{aa'}A^*{}_{,b}{e_d}^m\cD_mb_c\Big\}\sigma^a_{\alpha\dot\alpha}
\end{align}

The $T_\xi$-terms from \eqref{62} read in components
\begin{align}
\phone_\xi=&-2^{-2}\big(\cD^a\Phi-\cD^a\Phid\big)\big(\cD_a\Phi-\cD_a\Phid\big)\notag\\
&-2^{-4}\big(\cD^a\Phi\cD_a\Phi-\cD^a\Phid\cD_a\Phid\big)^2\\
\lc{\phone_\xi}_\tf=&-2^{-2}\big(A^{,m}-A^{*,m}\big)\big(A_{,m}-A^*{}_{,m}\big)+2^{-2}\sqrt2\big(A^{,m}-A^{*,m}\big)\big(\psi_m\chi-\bar\chi\bar\psi_m\big)\notag\\
&-2^{-4}\big(A^{,m}A_{,m}-A^{*,m}A^*{}_{,m}\big)^2\notag\\
&+2^{-3}\sqrt2\big(A^{,m}A_{,m}-A^{*,m}A^*{}_{,m}\big)\big(A^{,n}\psi_n\chi-A^{*,n}\bar\chi\bar\psi_n\big)\\
\cDla\phone_\xi=&-2^{-1}\big(\cDla\cD^a\Phi\cD_a\Phi+\cDla\cD^a\Phid\cD_a\Phid-\cDla\cD^a\Phi\cD_a\Phid-\cD^a\Phi\cDla\cD_a\Phid\big)\notag\\
&-2^{-2}\big(\cD^a\Phi\cD_a\Phi-\cD^a\Phid\cD_a\Phid\big)\big(\cDla\cD^a\Phi\cD_a\Phi-\cDla\cD^a\Phid\cD_a\Phid\big)\\
\lc{\cDla\phone_\xi}_\of=&-2^{-1}\big(\sqrt2\hD^a\chi_\a+\I\frac1{24}\sqrt2b_d(\sigma^d\bar\sigma^a\chi)_\a+\I\frac14\sqrt2b_a\chi_\a-\I\frac18\sqrt2b_d(\sigma_a\bar\sigma^d\chi)_\a\notag\\
&\qquad\qquad+\I\frac16\sqrt2(\sigma^a\bar\chi)_\a M^*\big)\big(A_{,a}-A^*{}_{,a}\big)\notag\\
&-2^{-2}\big(A^{,a}A_{,a}-A^{*,a}A^*{}_{,a}\big)\big([\sqrt2\hD^a\chi_\a+\I\frac1{24}\sqrt2b_d(\sigma^d\bar\sigma^a\chi)_\a\notag\\
&\hskip3em+\I\frac14\sqrt2b_a\chi_\a-\I\frac18\sqrt2b_d(\sigma^a\bar\sigma^d\chi)_\a]A_{,a}+\I\frac16\sqrt2(\sigma^a\bar\chi)_\a M^*A^*{}_{,a}\big)\\
\lc{\cDDa\phone_\xi}_\zf=&-2^{-1}\big(A^{,a}-A^{*,a}\big)\big(\lc{\cDua\cDla\cD_a\Phi}_\zf-\lc{\cDua\cDla\cD_a\Phid}_\zf\big)\notag\\
&-2^{-2}\big(A^{,a}A_{,a}-A^{*,a}A^*{}_{,a}\big)\big(\lc{\cDDa\cD^a\Phi\cD_a\Phi}_\zf-\lc{\cDDa\cD^a\Phid\cD_a\Phid}_\zf\big)\\
=&\phantom{+}2^{-1}\big(A^{,a}-A^{*,a}\big)\big(\I\frac83Fb_a+4F_{,a}+\frac23M^*[A_{,a}+A^*{}_{,a}]\big)\notag\\
&+2^{-2}\big(A^{,a}A_{,a}-A^{*,a}A^*{}_{,a}\big)\big(\I\frac83Fb^bA_{,b}+4F^{,b}A_{,b}+\frac23M^*[(\p A)^2+(\p A^*)^2]\big)
\end{align}

One can see explicitly see from the above expressions that the contributions of the $T_\phi$- and $T_\xi$-terms to the equation of motion of $b_m$ vanish in the ghost condensate background, where $\xi=\chi=\psi=M=F=0$.

\bibliographystyle{apsrev}
\bibliography{masterGC2}

\end{document}